\documentclass{article}
\usepackage{graphicx} 
\usepackage[margin=1in]{geometry}
\usepackage{graphicx} 
\usepackage{amsthm,amsfonts,amssymb,amsmath,comment,slashed,mathtools,enumitem,hyperref}
\usepackage{color}
\usepackage{tikz}
\allowdisplaybreaks

\newtheorem{theorem}{Theorem}[section]
\newtheorem{definition}{Definition}[section]
\newtheorem{proposition}{Proposition}[section]
\newtheorem{lemma}{Lemma}[section]
\newtheorem{remark}{Remark}[section]

\newcommand{\ml}{\left(}
\newcommand{\mr}{\right)}
\newcommand{\dv}{\partial_v}
\newcommand{\du}{\partial_u}

\newcommand{\D}{\mathrm{d}}

\title{\Large{A Scale-Critical $\mu<\frac38$ Extension Principle\\ for Spherically Symmetric Einstein--Scalar Field System}}
\author{Xinliang An$^1$, Haoyang Chen$^2$, Jaydeep Singh$^3$}

\date{}

\AtEndDocument{%
  \par
  \medskip
  \begin{tabular}{@{}l@{}}%
    \textsc{$^1$Department of Mathematics, National University of Singapore, Singapore}\\
    \textit{E-mail address}: \texttt{matax@nus.edu.sg}\\
     \textsc{$^2$School of Mathematical Sciences, Nankai University, China}\\
    \textit{E-mail address}: \texttt{chenhy@nankai.edu.cn}\\
    \textsc{$^3$School of Mathematics, Princeton University, USA}\\
    \textit{E-mail address}: \texttt{jaydeeps@math.princeton.edu}\\
  \end{tabular}}

\begin{document}

\maketitle
\begin{abstract}

We present a new quantitative, scale-critical $C^1$ extension principle for the spherically symmetric Einstein--scalar field system. We derive estimates for solutions in the causal past of a possibly singular endpoint $\mathcal{O}$ of the center.  Assume that there is no trapped surface in this region and assume that the Doppler exponent (blue shift) along each incoming null cone is uniformly bounded. We then prove that the solution can extend regularly across $\mathcal{O}$ provided the mass ratio $\mu=2m/r$ satisfies $\sup\mu<\frac38$. This improves the small-mass-ratio condition in Christodoulou's extension principle by an explicit threshold of order one. Our proof has three new ingredients. First, we derive several new uniform scale-critical pointwise estimates for the scalar field and its derivatives. Second, we obtain refined estimates in an exterior self-similar domain within the triangular causal past region. Third, we establish a family of higher-order Morawetz-type identities and identify the ranges of $\mu$, with which there are coercive spacetime estimates. The unweighted identity yields the threshold $\mu\leq\frac14$, while employing extra self-similar weight raises the threshold to $\mu<\frac38$ and provides a weighted control, which implies that the solution's BV norm along an outgoing null cone is small, so that Christodoulou's theory of bounded variation solutions gives the extension.

\end{abstract}

\tableofcontents

\section{Introduction}

Under spherical symmetry, we study the extension principle for the following Einstein--scalar field system:
\begin{equation}\label{ES}
\begin{cases}
\mbox{Ric}_{\mu\nu}-\frac{1}{2}Rg_{\mu\nu}=2T_{\mu\nu},\\
T_{\mu\nu}=\partial_{\mu}\phi \partial_{\nu}\phi-\frac{1}{2}g_{\mu\nu}\partial^{\sigma}\phi \partial_{\sigma}\phi. 
\end{cases}
\end{equation}

\noindent In the evolution, a spherically symmetric solution can lose regularity for the first time at the center $\Gamma$, where the area radius $r$ vanishes. A basic question is therefore which quantities, defined in the causal past of a possibly singular endpoint $\mathcal{O}$ of the center, determine whether the solution extends regularly beyond $\mathcal{O}$. In his analysis of gravitational collapse for this system \cite{christ1,christ2,christ3,christ4}, Christodoulou answered this question in terms of the dimensionless mass ratio $\mu=2m/r$, where $m$ is the Hawking mass. In \cite{christ2} he proved that if $\mu$ is sufficiently small in the causal past of a point on $\Gamma$, then the solution extends regularly to a full neighbourhood of that point. The threshold in this criterion remains to be improved, and the argument is perturbative in $\mu$. A second quantity which governs the behaviour near $\mathcal{O}$ is the blue shift along incoming null cones, measured by the Doppler exponent. It enters Christodoulou's analysis of the instability of naked singularities \cite{christ4}, and An--Tan \cite{antan} proposed an extension criterion in which the smallness of $\mu$ is replaced by the finite Doppler exponent.

\vspace{1mm}
The purpose of the present work is to improve the smallness condition on $\mu$ to the regime of order one. Assuming that there is no trapped sphere and that the Doppler exponent is uniformly bounded, we prove a $C^1$ extension principle up to the scale-critical threshold $\mu<\frac38$. Reaching an explicit threshold of order one is not a perturbative matter. The Hawking mass $m$, which represents the natural energy, is subcritical. The bound $2m=\mu r<r$ immediately excludes the concentration of $m$ as $r\to0$, but this gives no control of the scale-invariant quantity $\mu=2m/r$, which moreover has no monotonicity property. Christodoulou's analysis also shows that $\mu$ tends to zero along every fixed incoming null cone, but this does not exclude concentration of $\mu$ for example along a very steep timelike curve approaching $\mathcal{O}$. Our proof addresses these difficulties and gives quantitative estimates in three steps. We first derive uniform scale-critical pointwise estimates for the scalar field, assuming only the absence of trapped spheres and the finite Doppler exponent (blue shift). We then obtain improved estimates in an exterior self-similar region, showing that any remaining singular behaviour must concentrate at a scale strictly faster than the self-similar scale. Finally, we establish a family of higher-order Morawetz-type identities and identify the ranges of $\mu$ in which they are coercive. The unweighted identity excludes concentration of $\mu$ under $\mu\leq\frac14$, while a self-similar weight raises the threshold to $\mu<\frac38$ and provides the weighted control under which the solutions' BV norms become small, so that Christodoulou's small BV theory and extension theorem apply. 

\subsection{Main theorems}

Let $(\mathcal{Q},{}^{(4)}g,\phi)$ be a solution of the spherically symmetric Einstein--scalar field system in the causal past of a potentially singular endpoint $\mathcal{O}$ on the axis. We assume that $\mathcal{Q}$ admits a global double-null coordinate system $(u,v)$ in which
\[
    \mathcal{Q}=\{-1\leq u<0,\ u\leq v\leq0\}.
\]
The endpoint $\mathcal{O}$ is formally represented by $(u,v)=(0,0)$ but is not part of the spacetime, and the axis of symmetry is\footnote{We call this double-null coordinate system \emph{axis-normalized}. Another natural choice is the normalization $r(u,0)=|u|$, which we call \emph{horizon-normalized}.}
\[
    \Gamma=\{u=v,\ u<0\}.
\]
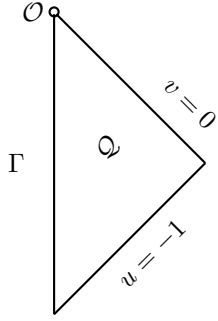
\begin{figure}[htbp]
	\centering
	\begin{tikzpicture}
		\begin{scope}[thick]
			\draw (0,0) -- (0,3.97);
			\draw (0,0) -- (2,2);
			\draw(0.03,3.97) -- (2,2);
			\node at (0,4)[circle,draw,inner sep=1.2pt]{};
			\node at (-0.5,2) {$\Gamma$};
			\node[anchor = east] at (0,4) {$\mathcal{O}$};
			\node[rotate=45] at (0.7,2.2) { $\mathcal{Q}$};
			\node[rotate=45] at (1.3,0.8) { ${u=-1}$};
			\node[rotate=-45] at (1.8,2.8) { $v=0$};
		\end{scope}
	\end{tikzpicture}
	\caption{Spacetime region $\mathcal{Q}$.}
	\label{domain}
\end{figure}

We impose the following assumptions:
\begin{enumerate}[label=(\Alph*)]
	\item There are no trapped or anti-trapped spheres, i.e. $\lambda > 0$, $\nu < 0$, and $1-\mu > 0$.
	\item On every outgoing cone $\Sigma_u$, $u\in[-1,0)$, the induced fields satisfy $r(u,\cdot),\phi(u,\cdot)\in C^2([u,0])$.
\end{enumerate}

\begin{definition}
	A solution satisfying \emph{(A)}--\emph{(B)} is called a $\mathbf{C^1}$-\textbf{admissible interior}.
\end{definition}
\begin{definition}
	A $C^1$-admissible interior satisfies the $\mathbf{C^1}$-\textbf{extension property} if, for every $C^2$ extension $r_0(-1,v),\phi_0(-1,v)$ of the outgoing data, there exists a sufficiently small $\delta>0$ and a $C^1$ extension of the interior spacetime to
	\[
	    \widetilde{\mathcal Q}=\{-1\leq u<\delta,\ u\leq v\leq\delta\}.
	\]
\end{definition}

For a fixed $\mu_{*} \in (0,1)$, we introduce the assumption
\begin{equation}
	\label{eq:extprin:assumpA_general} \tag{$\text{C}_{\mu_*}$} \sup_{\mathcal{Q}} \mu(u,v) \leq \mu_*.
\end{equation}
Christodoulou's extension theorem \cite{christ2} may then be formulated as follows.
\begin{theorem}[Christodoulou, \cite{christ2}]
	\label{thm1}
	There exists a sufficiently small $\varepsilon_0\in(0,1)$ such that every $C^1$-admissible interior satisfying $(\mathrm{C}_{\varepsilon_0})$ also satisfies the $C^1$-extension property.
\end{theorem}
\begin{remark}
	\label{rmk:ext_prin_ksquared_14}
	A natural question concerns the optimal value of $\mu_*$ in the above theorem. This value may depend on the precise regularity class of data, but we claim that even in the smooth class it must be strictly less than $\frac14$. This is a consequence of the explicit formulas for $\mu$ in the $k^2 = \frac13$ interior, {cf. \cite{christ3}}. The interior region of this spacetime is $C^\infty$ and satisfies $\mu \leq \frac14$ globally, but is \textit{not} extendible even in $\text{BV}$ due to $\mu(u,0) = \frac14 \not\rightarrow 0$.
\end{remark}

We prove a complementary result under the additional assumption of finite blueshift along ingoing null cones. More precisely, we measure the frequency shift by the Doppler exponent

\[
\gamma(v):=\int_{(-1,v)}^{(v,v)}\Big(\frac{\mu(-\nu)}{\rho r}\Big)(u',v) du'.
\]

We assume:
\begin{itemize}
	\item[(F)] There exists a constant $C>0$ such that, for every $v\in[-1,0)$,
	\begin{equation}\label{eq:finite_blueshift}
	\gamma(v) \leq C.
	\end{equation}
\end{itemize}

\begin{theorem}
	\label{thm2}
	Every $C^1$-admissible interior satisfying $(\mathrm{C}_{\frac14})$ and \emph{(F)} also satisfies the $C^1$-extension property.
\end{theorem}
The main result of this paper improves the threshold further by means of a self-similar multiplier.
\begin{theorem}
	\label{thm3}
	Let $\delta\in(0,\frac38)$. Every $C^1$-admissible interior satisfying $(\mathrm{C}_{\frac38-\delta})$ and \emph{(F)} also satisfies the $C^1$-extension property.
\end{theorem}
\begin{remark}
	In gravitational collapse, the finite-blueshift condition \emph{(F)} is a natural requirement for regular extension. When the Doppler exponent diverges, Christodoulou's instability mechanism for naked singularities \cite{christ4} generically leads to trapped-surface formation.
\end{remark}

\subsection{$C^1$ extension principle in Christodoulou's work}
For the spherically symmetric Einstein--scalar field system, Christodoulou proved the following theorem:

\noindent \textbf{Theorem (Christodoulou \cite{christ2}, $C^1$ extension theorem).} {\itshape Set $v_* > 0$ and assume that for every $\hat v \in (0,v_*)$ there is a $C^1$ solution on $\mathcal{D}(0,\hat v)$. There exists a constant $\varepsilon' > 0$ such that, if

\[
\lim_{u \rightarrow v_*} \sup_{\mathcal{D}(u,v_*)} \mu(u,v) < \varepsilon',
\]

	then there exist $\tilde{v} > v_*$ and a $C^1$ solution in the extended region $\mathcal{D}(0,\tilde{v})$.}

We briefly recall Christodoulou's argument. Define
\begin{equation*}
	\alpha:=\frac{\partial_v(r\phi)}{\partial_v r}, \quad \mbox{ and } \quad \alpha' := \frac{1}{\dv r} \dv \left( \frac{\dv (r\phi)}{\dv r} \right) =\frac{\dv \alpha}{\dv r}.
\end{equation*}
Differentiating gives
\begin{equation}\label{a' intro1}
	\begin{aligned}
		\du \alpha'
		=&\frac{-2\mu\partial_u r}{(1-\mu)r} \alpha' + \frac{-r\partial_u r}{1-\mu} \left(\frac{\dv \phi}{\dv r}\right)^3 + \frac{3\partial_u r\mu}{(1-\mu) r}\frac{\dv\phi}{\dv r}.
	\end{aligned}
\end{equation}

Integrating this equation gives
\begin{equation*}
	\begin{aligned}
		\alpha'(u,v)  &= \alpha'(0,v) +  \sum_{i=1}^{3} I_i(u,v),
	\end{aligned}
\end{equation*}
where $I_i$ is the integral of the $i$th term on the right-hand side of \eqref{a' intro1}.
This implies
\begin{equation}\label{alpha' Ii introduction}
	|\alpha'|(u,v)\leq \sup_{\{0\}\times [0, v_*]}|\alpha'|+\sum_{i=1}^{3}|I_i|(u, v).
\end{equation}

To prove Theorem \ref{thm1}, fix $0<\delta<1$ and set

\[
A'(u_1):=\sup_{\mathcal{D}(u_0, u_1, v_*)}(r^{\delta}|\alpha'|).
\]

Estimating each $I_i$ in \eqref{alpha' Ii introduction} gives

\[
|I_i|\lesssim \varepsilon' A'(u_1) r^{-\delta}.
\]

Since $\mu(u,v)$ is small in a region $\mathcal{D}(u_0, u_1, v_*)$ near $O'$, combining the contributions of these three terms yields
\[
A'(u_1)\lesssim 1.
\]

It follows that the solution extends to $\mathcal{D}(0,v_*)$.

A similar argument extends the $C^1$ solution beyond the triangular boundary. Consider the thin rectangular region
\begin{equation*}
	U_{\xi}(u_1) = [u_0,u_1) \times [v_*,v_*+\xi].
\end{equation*}
We derive uniform bounds for the quantities

\[
\overline{A'_{\xi}}(u_1) := \sup_{U_{\xi}(u_1)} (r^{\delta}|\alpha'|) \quad
\mbox{ and }
\quad A'_* := \sup_{u_1\in(u_0,v_*)}A'(u_1).
\]

For $0<\delta<1$, we obtain

\[
\sup_{\mathcal{D}(u_0,u_1,v_*+\xi)}(r^{\delta}|\alpha'|) \leq \max\{ A'(u_1), \overline{A'_\xi}(u_1)\} \leq \max\{A'_*, \overline{A'_\xi}(u_1)\}.
\]

By replacing $A'(u_1)$ with $\max\{A'_*, \overline{A'_\xi}(u_1)\}$, we also obtain an analogous bound for $\overline{A'_{\xi}}(u_1)$. As above, improved estimates for $m(u,v)$ and $\mu(u,v)$ then give a uniform bound for $|\alpha'(u,v)|$ in $U_{\xi}(u_1)$, which yields the desired extension.

\subsection{Main difficulties}
The main difficulties stem from the scale-critical nature of the extension criterion. We highlight three obstacles that make the bounds $\mu\leq\frac14$ and $\mu<\frac38$ non-trivial.
\begin{itemize}
	\item Christodoulou's analysis shows that $\mu$ tends to zero along every fixed incoming null cone $\{v=\mathrm{const.}\}$. This does not, however, exclude concentration along a moving curve approaching $\mathcal O$, for instance a timelike curve. We first rule out concentration at the self-similar scale and then use a Morawetz-type estimate to exclude concentration at faster scales under the stated bounds on $\mu$.

	\begin{figure}[htbp]
		\centering
	\tikzset{every picture/.style={line width=0.75pt}} 

	\begin{tikzpicture}[x=0.75pt,y=0.75pt,yscale=-1,xscale=1]

		\draw [fill={rgb, 255:red, 155; green, 155; blue, 155 }  ,fill opacity=0.64 ]   (170.65,49.79) -- (276.71,155.85) ;
		\draw [shift={(170.65,49.79)}, rotate = 45] [color={rgb, 255:red, 0; green, 0; blue, 0 }  ][line width=0.75]      (0, 0) circle [x radius= 3.35, y radius= 3.35]   ;
		\draw [fill={rgb, 255:red, 155; green, 155; blue, 155 }  ,fill opacity=0.64 ]   (170.65,49.79) -- (170.65,261.91) ;
		\draw [fill={rgb, 255:red, 155; green, 155; blue, 155 }  ,fill opacity=0.64 ]   (170.65,261.91) -- (276.71,155.85) ;
		\draw    (170.65,49.79) .. controls (179,114) and (177,187) .. (209,224) ;

		\draw (148,142.4) node [anchor=north west][inner sep=0.75pt]    {$\Gamma $};
		\draw (149,31.4) node [anchor=north west][inner sep=0.75pt]    {$\mathcal{O}$};
		\draw (186,135.4) node [anchor=north west][inner sep=0.75pt]    {$\Gamma _{\frac{3}{8}}$};

	\end{tikzpicture}
		\caption{Possible concentration along a timelike curve $\Gamma_{\frac38}$ where $\mu=\frac38$.}
	\end{figure}
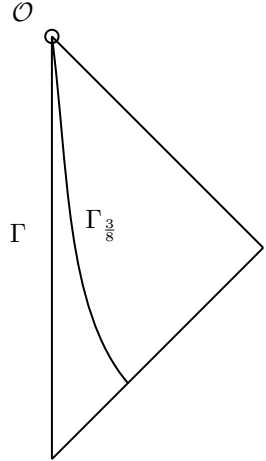
	\item The mass ratio $\mu$ is scale invariant, whereas the Hawking mass $m$, which represents the natural energy, is subcritical. Indeed, $2m=\mu r<r$ immediately implies non-concentration of $m$ as $r\to0$, but this provides no corresponding control of $\mu=2m/r$. Moreover, unlike $m$, the mass ratio has no useful monotonicity property. Thus neither scaling nor ordinary energy non-concentration suffices to prove a scale-critical extension criterion. The thresholds $\mu\leq\frac14$ and $\mu<\frac38$ obtained here require additional structure.

	\item Uniform $L^\infty$ estimates for scale-critical quantities such as $rD_v\phi$ are essential but delicate. At this stage, the only geometric restriction on $\mu$ is the no-trapped-surface condition $\mu<1$. We derive these estimates from this condition and the finite-blueshift assumption.
\end{itemize}

\subsection{Proof strategy and new ingredients}
We now outline the proof, explaining how the pointwise, self-similar, and energy estimates combine to yield the extension theorems. We summarize the new ingredients in the next subsection.

\subsubsection{Uniform pointwise estimates}
In Section \ref{subsec:extprin_selfsimilarregion}, we derive pointwise estimates using the no-trapped-surface condition (A) and the finite-blueshift condition (F).

We first establish a local estimate for the variation of $r\partial_v \phi$. Fix a point $P=(u_0,v_0) \in \mathcal{Q}$, and let $C_{P, \delta} = \{(u',v'): u' \in [-1,u_0], v' \in [v_0-\delta, v_0] \}$ denote a characteristic rectangle of $v$-width $\delta$ extending back to the initial data surface $\{u=-1\}$. A bootstrap argument gives the following local control; see Lemma \ref{lem:local} and its proof:
\begin{lemma}
	Let $r_0 = r(u_0,v_0)$, and define
	\begin{equation*}
		F(P) = 1 + \sup_{u \in [-1,u_0]}|r(u,v_0)D_v\phi(u,v_0)|.
	\end{equation*}
	For all sufficiently small $\delta'>0$, the following bounds hold in $C_{P, \delta' F(P)^{-1} r_0}$:
	\begin{equation}
		|r \partial_v \phi(u,v) - r\partial_v \phi(u,v_0)| \lesssim \delta' F(P),
	\end{equation}
	and
		\begin{equation}
		\sup_{C_{P,\delta}}|r \partial_v \phi| \leq C F(P),
	\end{equation}
	where $\delta \doteq \delta' F(P)^{-1} r_0$ is sufficiently small.
\end{lemma}

This lemma yields the following uniform bounds:
	\begin{align}
		& \sup_{\mathcal{Q}}(|\phi| + |rD_v\phi| + |rD_u\phi| + |r^2D_{vv}^2\phi|) \leq C, \\
		& |rD_v\phi| \lesssim \mu^{\frac13}.
	\end{align}

	To prove that $rD_v\phi$ is globally bounded, suppose instead that there is a sequence $p_i=(u_i,v_i)$ with $u_i$ increasing, $u_i \rightarrow 0$, such that $c_i \doteq |rD_v\phi(u_i,v_i)| \rightarrow \infty$.

	Set $r_i \doteq r(u_i,v_i)$. The local estimate above gives a sufficiently small $\delta'>0$, independent of $i$, such that
	\begin{equation*}
		|r\partial_v\phi(u_i,v)-r\partial_v\phi(u_i,v_i)| \leq \frac12 c_i
	\end{equation*}
	holds for $v \in I_i \doteq [v_i - \delta' c_i^{-1}r_i, v_i]$. In particular, we have the lower bound $|r\partial_v\phi(u,v)| \geq \frac12 c_i$ in $I_i$. Computing the Hawking mass in this interval gives
	\begin{align*}
		m(u_i,v_i) &\geq m(u_i,v_i)-m(u_i,v_i-\delta' c_i^{-1}r_i) \\
		&\gtrsim \int_{v_i-\delta' c_i^{-1}r_i}^{v_i} (r\partial_v\phi)^2(u_i,v')dv'\\
		&\gtrsim \delta' c_i r_i.
	\end{align*}
	This would imply
	\begin{align*}
		\mu(u_i,v_i) = \frac{2m(u_i,v_i)}{r_i} \gtrsim \delta' c_i \rightarrow \infty,
	\end{align*}
	a contradiction.

	Therefore $r\partial_v\phi$ is uniformly bounded. Moreover,
	\begin{align*}
		\phi(u,v)-\phi(-1,v) &= \int_{-1}^{u} \partial_u\phi(u',v) du' \\
		&= \int_{-1}^{u} \frac{(-\nu)\mu}{\rho r} (rD_v\phi)(u',v)du' + \big( (rD_v\phi)(-1,v)-(rD_v\phi)(u,v) \big),
	\end{align*}
	and the right-hand side is bounded by the estimates for $rD_v\phi$ and the Doppler integral.

	We next integrate the wave equation
	\begin{align*}
		\partial_v \partial_u (r\phi) = \frac{\mu \lambda \nu}{\rho r}\phi,
	\end{align*}
	from the axis to obtain $|rD_u\phi| \lesssim 1$.

	To refine the bound for $rD_v\phi$ in terms of the local value of $\mu$, fix $P=(u,v)$ and write $|r\partial_v \phi|(u,v) = \kappa$, where $\kappa\geq0$. We return to the local argument above and use \eqref{eq:stabilityest}. For sufficiently small $\delta_0$, there exists an interval $I = [v-\delta' F(p)^{-1} r(u,v), v]$ in the $v$-coordinate on which
	\begin{align} \label{discuss kappa intro}
		|r\partial_v \phi|(u,v') &\geq \kappa-C\delta' F(P)\\ \nonumber
		&\geq \kappa-\frac14 F(P),
	\end{align}
	where $C$ is the uniform constant in \eqref{eq:stabilityest}. Since $rD_v \phi$ is bounded, $F(P)$ is bounded by a fixed large constant $F$. For sufficiently small $\kappa<\frac12 F$, the mass estimate gives
	\begin{align*}
		m(u,v) &\gtrsim \int_{v-\delta' F(P)^{-1} r(u,v)}^{v}(r\partial_v\phi)^2\\
		&\gtrsim F^{-2} \kappa^3 r(u,v).
	\end{align*}
	It follows that $\mu \gtrsim \kappa^3 = |r\partial_v\phi|^3$.

	Finally, we estimate $r^2 D_{vv}^2\phi$. In view of the bound for $rD_v\phi$, it suffices to show that $|r\alpha'| \lesssim 1$, where $\alpha' = D_{vv}^2(r\phi)$.

	We compute
	\begin{align*}
		\partial_u(\lambda^2  \alpha') =  \frac{(-\nu)}{\lambda \rho} r (\partial_v\phi)^3 + \frac{3\nu \mu \lambda}{\rho r} (\partial_v\phi).
	\end{align*}
	Integrating from the initial data and using the bound on the Doppler integral gives the desired estimate.

These estimates also yield decay along the final cone $v=0$:
	\begin{align*}
		&\lim_{u \nearrow 0} \mu(u,0)=0,\\
		&\lim_{u \nearrow 0} (rD_v\phi)(u,0)=0.
	\end{align*}

\subsubsection{Refined estimates in the self-similar region}
In Section \ref{sec:selfsimilar}, we derive quantitative estimates in the exterior self-similar region. Define the \emph{self-similar coordinate} by

\[
x\triangleq \frac{r}{-u}.
\]

For fixed $0<\delta \ll 1$, define the \emph{exterior self-similar region}
\[
\mathcal{C}_\delta =\mathcal{Q} \cap \{x \geq \delta \},
\]

and the \emph{interior self-similar region}

\[
\mathcal{B}_\delta =\mathcal{Q} \cap \{0\leq x \leq \delta \}.
\]

Moreover, let

\[
\Gamma_\delta = \{x = \delta \}
\]
 denote their common timelike boundary.

\begin{figure}[htbp]
	\centering
	\tikzset{every picture/.style={line width=0.75pt}} 

	\begin{tikzpicture}[x=0.75pt,y=0.75pt,yscale=-1,xscale=1]

		\draw [fill={rgb, 255:red, 155; green, 155; blue, 155 }  ,fill opacity=0.64 ]   (172.31,51.45) -- (276.71,155.85) ;
		\draw [shift={(170.65,49.79)}, rotate = 45] [color={rgb, 255:red, 0; green, 0; blue, 0 }  ][line width=0.75]      (0, 0) circle [x radius= 3.35, y radius= 3.35]   ;
		\draw [fill={rgb, 255:red, 155; green, 155; blue, 155 }  ,fill opacity=0.64 ]   (170.65,49.79) -- (170.65,261.91) ;
		\draw [fill={rgb, 255:red, 155; green, 155; blue, 155 }  ,fill opacity=0.64 ]   (170.65,261.91) -- (276.71,155.85) ;
		\draw    (170.65,49.79) -- (208,224) ;

		\draw (148,142.4) node [anchor=north west][inner sep=0.75pt]    {$\Gamma $};
		\draw (149,31.4) node [anchor=north west][inner sep=0.75pt]    {$\mathcal{O}$};
		\draw (212,129.4) node [anchor=north west][inner sep=0.75pt]    {$\mathcal{C}_{\delta }$};
		\draw (183,89.4) node [anchor=north west][inner sep=0.75pt]    {$\Gamma _{\delta }$};
		\draw (173,164.4) node [anchor=north west][inner sep=0.75pt]    {$\mathcal{B}_{\delta }$};

	\end{tikzpicture}
\end{figure}
The uniform $L^\infty$ estimates and the boundary behavior along $\{v=0\}$ yield sharp bounds in the exterior self-similar region:
	\begin{align}
		\lim_{u \rightarrow 0}\sup_{\mathcal{C}_{\delta}\cap\{u' \geq u\}}\mu(u',v) &= 0, \\
		\sup_{\mathcal{C}_{\delta}} |r\partial_v\phi| &\lesssim_\delta 1,\\
		\sup_{\mathcal{C}_{\delta}} |\phi| &\lesssim_\delta 1.
	\end{align}

The same monotonicity properties yield the integrated bounds
	\begin{align}
		\int_{\Gamma_\delta} \frac{(-\nu) \mu}{\rho r} du + \iint_{\mathcal{C}_{\delta}} \Big(\frac{m}{r^3} + (\partial_v\phi)^2\Big) (-\nu)dudv  \lesssim_\delta 1.
	\end{align}

\subsubsection{Morawetz-type higher-order energy estimates}
In Section \ref{subsec:extprin_multipliers}, we derive higher-order weighted energy identities and use them to establish Morawetz-type estimates.

The weighted identity takes the form
	\begin{align*}
		\partial_u \Theta + G_1(D_{vv}^2\phi)^2 + G_2(D_v\phi)^6 + G_3(D_v\phi)^4 + G_4(D_v\phi)^2 = \partial_v B ,
	\end{align*}
	where the coefficients and boundary flux depend on a suitably regular weight $\tilde{\Gamma}(u,v)$.

Choosing $\tilde{\Gamma}(u,v)= 1$ and $\tilde{\Gamma}(u,v)= x^\frac45$, respectively, and integrating the identity yields the Morawetz-type spacetime estimates
\begin{align}
	&\int_{\Sigma_{u}} r \lambda (\alpha')^2(u,v) dv+\iint\limits_{R(-1,u)} G_1 (D_{vv}^2\phi)^2 + G_2(D_v\phi)^6 + G_3(D_v\phi)^4 + G_4(D_v\phi)^2 dudv\\ \nonumber
	&+\int_{\underline{\Sigma}_0(-1,u)}\frac{r^3 \nu}{2 \rho}  (D_v\phi)^4  + \frac{-\nu r(2-5\mu)}{\rho}(D_v\phi)^2 du=   \int_{\Sigma_{-1}} r \lambda (\alpha')^2(-1,v) dv,
\end{align}
and
\begin{align}
	&\int_{\Sigma_{u}} r \lambda x^\sigma (\alpha')^2(u,v) dv+\iint\limits_{R(-1,u)} G_1 (D_{vv}^2\phi)^2 + G_2(D_v\phi)^6 + G_3(D_v\phi)^4 + G_4(D_v\phi)^2 dudv\\ \nonumber
	&-\int_{\underline{\Sigma}_0(-1,u)}B_0(u) du=   \int_{\Sigma_{-1}} r \lambda x^\sigma (\alpha')^2(-1,v) dv.
\end{align}
An analysis of the coefficient signs gives coercivity under the bounds $\mu\leq\frac14$ and $\mu < \frac38$, respectively.

\subsubsection{Non-concentration of the mass ratio for $\mu\leq\frac14$}
In Section \ref{mu less 1/4}, we rule out concentration of the mass ratio $\mu(u,v)$ near $\mathcal{O}$ and thereby prove Theorem \ref{thm2}. More precisely, we show that there cannot exist $\epsilon >0$ and a sequence $(u_i,v_i) \rightarrow 0$, with $u_i$ increasing and $u_i \rightarrow 0$, such that $\mu(u_i,v_i) \geq \epsilon$ for all $i$.

Assume that $\mu\leq \frac14$ in $\mathcal{Q}$ and that such a sequence concentrates at a scale strictly faster than the self-similar scale. To rule out this concentration, we use the higher-order energy estimate for $\tilde{\Gamma}=1$ to obtain
\begin{align*}
	\iint_{R(u_1,u_2)}(\partial_v\phi)^2 (-\nu) dudv \lesssim 1 .
\end{align*}
A related bulk integral involving the Hawking mass can be controlled by \eqref{eq:temp6}:
\begin{align*}
	\iint_{R(u_1,u_2)}\frac{2\lambda(-\nu)\mu}{\rho r^2}dudv \lesssim 1.
\end{align*}

Return to the sequence $(u_i,v_i)$ with $\mu(u_i,v_i) \geq \epsilon$. Let $r_i = r(u_i,v_i)$, and let $u_{*,i}, v_{*,i}$ be such that $r(u_i,v_{*,i}) - r_i = r(u_{*,i},v_i) - r_i = r_i$ holds. Passing to a subsequence, we may arrange that the characteristic regions $\mathcal{R}_i \doteq [u_{*,i},u_i] \times [v_i, v_{*,i}]$ are pairwise disjoint. Suppose $(u_{i_k}, v_{i_k})$ have been chosen, $1 \leq k \leq N$, subject to the condition that $r_{i_k}$ are strictly decreasing in $k$. By construction, we have $\mu \Big|_{\mathcal{R}_i} \gtrsim \epsilon$. This, however, forces the spacetime energy to diverge:
\begin{align*}
	\iint_{\mathcal{Q}}\frac{m}{r^3}(-\nu)dudv &\geq \sum_{i=1}^{\infty}\iint_{\mathcal{R}_i}\frac{m}{r^3}(-\nu)dudv \\
	&\gtrsim  \epsilon  \sum_{i=1}^{\infty} 1 \\
	&=\infty.
\end{align*}
This contradicts the higher-order energy estimate and rules out concentration of $\mu$ at the endpoint $\mathcal{O}$.

\subsubsection{Non-concentration of BV norms for $\mu<\frac38$}
In Section \ref{sec:mu less 3/8}, we prove Theorem \ref{thm3} by combining the self-similar estimates of Section \ref{sec:selfsimilar} with the weighted energy identities of Section \ref{subsec:extprin_multipliers}. This yields non-concentration of the relevant BV norm near $\mathcal{O}$.

\subsection{New ingredients}
The proof introduces three ingredients that may be useful beyond the present extension problem:
\begin{itemize}
	\item We derive non-trivial uniform, scale-critical pointwise estimates for the scalar field using only the no-trapped-surface and finite-blueshift assumptions.
	\item We obtain refined estimates in exterior self-similar regions, showing that any remaining singular behavior must concentrate at a scale strictly faster than the self-similar scale.
	\item We establish a family of higher-order Morawetz-type identities and identify their sharp coercivity ranges. The unweighted identity yields the threshold $\mu\leq\frac14$, while the self-similar weight $x^{4/5}$ raises the threshold to $\mu<\frac38$ and provides the weighted control needed for the BV argument.
\end{itemize}

\subsection{Related work in spherical symmetry}
We briefly review related work in spherical symmetry.
Christodoulou's foundational work \cite{christ1,christ2,christ3,christ4} established the framework for studying gravitational collapse in the Einstein--scalar field system. Of particular relevance here are his theory of solutions of bounded variation and the extension principle based on a sufficiently small mass ratio \cite{christ2}, which is a key ingredient in his instability argument \cite{christ4}. For extension principles and global structure, see also Dafermos \cite{MD05b}. The present work concerns the uncharged system, under the no-trapped-surface condition $\mu<1$ and the assumption of finite blueshift.

Self-similar naked singularities provide examples in which to study the blueshift mechanism. Following Christodoulou's construction \cite{christ3}, Singh constructed approximately self-similar naked-singularity spacetimes \cite{Singh} and studied the regularity dependence of blueshift for linear waves in their interiors \cite{Singh2}. In \cite{CicortasKehle26}, Cicortas--Kehle also constructed discretely self-similar exterior-naked singularities. 

An--Tan \cite{antan} studied weak cosmic censorship for the Einstein--Maxwell--charged scalar field system using double-null coordinates. On the instability of naked singularities for the Einstein-scalar field system, Liu--Li \cite{LL1} gave a direct proof using a priori estimates. A related line of work concerns the formation of trapped surfaces. Extensions of Christodoulou's trapped-surface criterion \cite{christ1} include work by An--Lim \cite{anlim} for charged scalar fields, Costa \cite{costacqg} for scalar fields with a positive cosmological constant.
\subsection{Organization of the paper}
The rest of the paper is structured as follows:

\begin{itemize}
	\item In Section \ref{pre}, we introduce the double-null gauge and derive the governing equations under spherical symmetry.
	
	\item In Section \ref{subsec:extprin_selfsimilarregion}, we establish pointwise estimates from the no-trapped-surface and finite-blueshift conditions.
	
	\item In Section \ref{sec:selfsimilar}, we prove refined estimates in the exterior self-similar region.
	
	\item In Section \ref{subsec:extprin_multipliers}, we derive higher-order Morawetz-type energy estimates and analyze their coercivity for different weights and bounds on $\mu$.

	\item In Section \ref{mu less 1/4}, we prove Theorem \ref{thm2} by ruling out concentration of $\mu$ under the assumption $\mu\leq\frac14$.
	
	\item {In Section \ref{sec:mu less 3/8}, we prove Theorem \ref{thm3} by obtaining small BV variation under the assumption $\mu<\frac38$ and applying Christodoulou's small-BV theory and extension theorem.}
\end{itemize}

\subsection{Acknowledgments}
Xinliang An is supported by MOE Tier 1 grant A-8002933-00-00. Haoyang Chen is supported by the Fundamental Research Funds for the Central Universities (050-63263077). 

\section{Preliminaries} \label{pre}
We first fix the double-null gauge and introduce the geometric quantities used throughout the paper. We then record the field equations and the identities needed for the estimates, before defining the spacetime regions over which we integrate.
\subsection{Spherically symmetric spacetimes}
Under spherical symmetry, the Lorentzian metric takes the form
\begin{equation*}
	g=-\Omega^2(u,v)dudv+r^2(u,v)(d\theta^2+\sin^2\theta d\varphi^2),
\end{equation*}
in coordinates $(u,v,\theta,\varphi)$, where $(u,v)$ are double-null coordinates on the quotient spacetime. Here $\Omega$ is the null lapse and $r$ is the area-radius function.

We fix the double-null gauge by imposing
\[
    \Gamma=\{u=v,\ u<0\},
    \qquad
    r(-1,v)=\frac12(v+1),
\]
where $\{u=-1\}$ is the initial outgoing cone.
The Hawking mass of each symmetry sphere is defined by
\begin{equation}\label{Hawking mass}
	m(u,v):=\frac{r}{2}(1+4\Omega^{-2}\partial_u r \partial_v r).
\end{equation}
The corresponding dimensionless mass ratio is
\[
    \mu:=\frac{2m}{r},
\]
so that 
\begin{equation*}
	1 - \frac{2m}{r} = g(\nabla r, \nabla r) = - 4\Omega^{-2}\partial_u r\partial_v r.
\end{equation*}

We also write
\begin{equation*}
	\nu = \partial_u r, \quad \lambda = \partial_v r, \quad \rho = 1-\mu.
\end{equation*}

Under the no-anti-trapped-surface condition $\nu<0$, we distinguish three regions of the quotient spacetime:

\begin{itemize}
	\item $(u,v)\in\mathcal{T}$ if and only if $\lambda(u,v)<0$, equivalently $\rho(u,v)<0$ or $\mu(u,v)>1$;
	\item $(u,v)\in\mathcal{R}$ if and only if $\lambda(u,v)>0$, equivalently $\rho(u,v)>0$ or $\mu(u,v)<1$;
	\item $(u,v)\in\mathcal{A}$ if and only if $\lambda(u,v)=0$, equivalently $\rho(u,v)=0$ or $\mu(u,v)=1$.
\end{itemize}
Thus $\mathcal{T}$, $\mathcal{R}$, and $\mathcal{A}$ denote the trapped region, regular region, and apparent horizon, respectively. Throughout the paper, we work in the regular region, where $\lambda>0$ and $\rho>0$ everywhere in $\mathcal Q$.

\subsection{Equations of motion}
In these coordinates, the Einstein--scalar field system \eqref{ES} takes the form
\begin{align}
	\partial_u\partial_v r&=\partial_u \lambda = \partial_v \nu = \frac{\mu \lambda \nu}{\rho r}, \label{eq:SSESF1}\\
	\partial_u\partial_v\phi &= -\frac{\lambda}{r}\partial_u\phi - \frac{\nu}{r}\partial_v \phi, \label{eq:SSESF6}\\
\partial_u\partial_v\log \Omega^2&=-2\partial_u\phi\partial_v\phi+\frac{1}{2}r^{-2}\Omega^2+2r^{-2}\partial_ur \partial_v r, \label{propeq:Omega}\\
    \partial_v (\nu^{-1}\rho) &=-4 \partial_v(\Omega^{-2}\lambda) = -\frac{\rho}{\lambda\nu}r(\partial_v\phi)^2, \label{eq:EP:SSESF14}\\
\partial_u (\lambda^{-1}\rho) &=-4 \partial_u(\Omega^{-2}\nu) = -\frac{\rho}{\lambda\nu}r(\partial_u\phi)^2. \label{eq:EP:SSESF15}
\end{align}
The scalar wave equation can also be written in transport form:
\begin{align}
	\partial_u(r\partial_v\phi) &= -\lambda \partial_u\phi, \label{eq:SSESF7}\\ 
	\partial_v(r\partial_u\phi) &= -\nu \partial_v\phi. \label{eq:SSESF8}
\end{align}
The Hawking mass and mass ratio satisfy the transport equations
\begin{align}
	\partial_u m &= \frac{\rho}{2\nu}(r\partial_u\phi)^2, \label{eq:SSESF2}\\ 
	\partial_v m &= \frac{\rho}{2\lambda}(r\partial_v\phi)^2, \label{eq:SSESF3}\\
	\partial_u \mu &= -\frac{\nu}{r}\mu + \frac{\rho}{\nu}r(\partial_u\phi)^2, \label{eq:SSESF4}\\
	\partial_v \mu &= -\frac{\lambda}{r}\mu + \frac{\rho}{\lambda}r(\partial_v\phi)^2. \label{eq:SSESF5}
\end{align}
For the pointwise estimates, it is convenient to use the rescaled scalar-field quantities
\begin{equation*}
	\psi = r\phi, \quad r D_v \phi, \quad r D_u \phi,
\end{equation*}
where
\[
    D_v=\lambda^{-1}\partial_v,
    \qquad
    D_u=(-\nu)^{-1}\partial_u
\]
are gauge-invariant derivatives. The equations above give
\begin{align}
	\partial_u(r D_v\phi) &= -\frac{\mu \nu}{\rho r}(r D_v\phi) - \partial_u\phi \label{eq:SSESF9}\\
	\partial_v(r D_u\phi) &= -\frac{\mu \lambda}{\rho r}(r D_u\phi) + \partial_v\phi \label{eq:SSESF10}\\
	\partial_u \partial_v \psi &= \frac{\mu \lambda \nu}{\rho r^2}\psi \label{eq:SSESF11}
\end{align}
For the higher-order estimates, we introduce
\begin{equation*}
 \alpha'=D_{vv}^2(r\phi), \quad \underline{\alpha}' = D_u^2(r\phi),
\end{equation*}
which gives the relation
\begin{equation} \label{alpha' relation}
	\alpha' = r D_{vv}^2\phi + 2 D_v\phi.
\end{equation}
The outgoing higher-order quantity also satisfies the transport equation
\begin{equation} \label{eq:SSESF18}
	\partial_u \alpha' = \frac{2\mu(-\nu)}{\rho} D_{vv}^2\phi + \frac{(-\nu)}{\rho}r (D_v\phi)^3 + \frac{(-\nu)\mu}{\rho r}D_v\phi. 
\end{equation}
\subsection{Spacetime regions}
We work on the triangular domain
\[
    \mathcal Q=\{-1\leq u<0,\ u\leq v\leq0\},
\]
whose endpoint $\mathcal O$, formally represented by $(0,0)$, is not part of the spacetime.

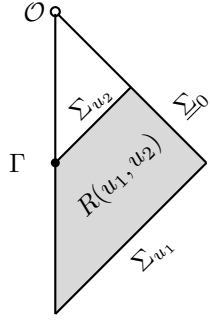
\begin{figure}[htbp]
	\centering
	\begin{tikzpicture}
		\begin{scope}[thick]
			\fill[gray!30] (0,0) to (0,2) to (1,3) to (2,2) to (0,0);
			\draw (0,0) -- (0,3.97); 
			\draw (0,0) -- (2,2);
			\draw(0.03,3.97) -- (2,2);
			\node at (0,2)[circle,fill,inner sep=1.2pt]{};
			\node at (0,4)[circle,draw,inner sep=1.2pt]{};
			\node at (-0.5,2) {$\Gamma$};
			\node[anchor = east] at (0,4) {$\mathcal{O}$};
			\node[rotate=45] at (0.48,2.8) { $\Sigma_{u_2}$};
				\node[rotate=45] at (1.3,0.8) { $\Sigma_{u_1}$};
				\node[rotate=45] at (1.8,2.8) { $\underline{\Sigma}_0$};
				\node[rotate=45] at (0.85,1.83) { $R(u_1,u_2)$};
			\draw (0,2) -- (1,3);
		\end{scope}
	\end{tikzpicture}
	\caption{The spacetime region $R(u_1,u_2)$.}
	\label{FETFig1}
\end{figure}

Let $\Sigma_u$ and $\underline{\Sigma}_v$ denote the outgoing and ingoing null cones of constant $u$ and $v$, respectively. For $u_1\leq u_2<0$, we define the region between two outgoing cones by
\[
    R(u_1,u_2)=\mathcal Q\cap\{u_1\leq u\leq u_2\}.
\]
The boundary of this region consists of the axis segment $\Gamma(u_1,u_2)$, the outgoing cones $\Sigma_{u_1}$ and $\Sigma_{u_2}$, and the final ingoing segment $\underline{\Sigma}_0(u_1,u_2)$; see Figure \ref{FETFig1}.

\section{Pointwise estimates} \label{subsec:extprin_selfsimilarregion}
In this section, we derive uniform scale-critical pointwise estimates from the no-trapped-surface condition \emph{(A)} and the finite-blueshift condition \emph{(F)}. At this stage, we use only the geometric bound $\mu<1$.

We first derive geometric estimates from the finite-blueshift assumption \emph{(F)} and the monotonicity properties of the system.

\begin{lemma}
	\label{lem:extprin_1}
	The following estimates hold:
	\begin{align}
		\sup_{\mathcal{Q}}|\log \lambda| &\lesssim 1,  \label{eq:3.1}\\
		\sup_{\mathcal{Q}}|\log (1-\mu)| &\lesssim 1. \label{eq:3.2}
	\end{align}
	There is also a constant $c>0$ such that, for all $(u,v)\in\mathcal{Q}$,
	\begin{align}
		(-\nu)(u,v) \geq c > 0, \label{eq:3.3}
	\end{align}
	and constants $c_1, c_2 >0$ such that 
	\begin{align}
		c_1(v-u) \leq r(u,v) \leq c_2 (v-u). \label{eq:3.4}
	\end{align}
\end{lemma}
\begin{proof}
	To prove \eqref{eq:3.1}, rewrite \eqref{eq:SSESF1} as an equation for $\log\lambda$, integrate in $u$ from $\{u=-1\}$, and apply assumption \emph{(F)}.
	
	Next, the positivity of $\mu$ gives $1-\mu \leq 1$. In particular, $\log(1-\mu) \leq 0$.
	
	By \eqref{eq:EP:SSESF15}, $\log(\rho^{-1}\lambda)$ is decreasing in $u$. Comparison with the initial data therefore gives $\log(\rho^{-1}\lambda) \leq C$ for a constant $C>0$. Equivalently, $\log (1-\mu) \geq -C + \log\lambda \geq C'$ for a finite constant $C'$, which proves \eqref{eq:3.2}.
	
	The gauge normalization and the bound on $\lambda$ give $(-\nu)\big|_\Gamma \gtrsim 1$. By \eqref{eq:SSESF1}, $-\nu$ is increasing in $v$, which yields the uniform lower bound \eqref{eq:3.3}.
	
	Finally, to prove \eqref{eq:3.4}, integrate $\partial_v r = \lambda$ from the center and use \eqref{eq:3.1}. For example, 
	\begin{align*}
		r(u,v) &= r(u,u) + \int_{u}^{v}\lambda(u,v')dv' \\
		&\leq c_2(v-u)
	\end{align*}
	for an appropriate constant $c_2$. An identical argument gives the lower bound.
\end{proof}

We next establish a local stability estimate for $r\partial_v\phi$. Fix $P=(u_0,v_0)\in\mathcal Q$ and define
\[
    C_{P,\delta}=\{(u',v'):-1\leq u'\leq u_0,\ v_0-\delta\leq v'\leq v_0\}.
\]
Thus $C_{P,\delta}$ is a characteristic rectangle of $v$-width $\delta$ extending back to the initial cone $\{u=-1\}$.
\begin{lemma} \label{lem:local}
   Let $r_0=r(u_0,v_0)$ and define
    \begin{equation*}
        F(P) = 1 + \sup_{u \in [-1,u_0]}|r(u,v_0)D_v\phi(u,v_0)|.
    \end{equation*}
    There exists a uniform constant $\delta_0>0$ such that, for every $0<\delta'\leq\delta_0$, throughout $C_{P,\delta'F(P)^{-1}r_0}$ we have
    \begin{equation}
    \label{eq:stabilityest}
        |r \partial_v \phi(u,v) - r\partial_v \phi(u,v_0)| \lesssim \delta' F(P).
    \end{equation}
\end{lemma}

\begin{proof}
    Fix a large constant $C\gg1$ and introduce the bootstrap assumption\footnote{The argument is adapted from Liu--Li \cite{LL1}, who used a similar bootstrap to control the Einstein--scalar field system in the exterior of naked-singularity spacetimes.}
    \begin{equation}
    \label{eq:appboot}\tag{A1}
    \sup_{C_{P,\delta}}|r \partial_v \phi| \leq C F(P),
\end{equation}
where $\delta\doteq\delta'F(P)^{-1}r_0$.
	\begin{figure}[htbp]
	\centering
	\tikzset{every picture/.style={line width=0.75pt}} 
	
	\begin{tikzpicture}[x=0.75pt,y=0.75pt,yscale=-1,xscale=1]
		
		\draw [fill={rgb, 255:red, 155; green, 155; blue, 155 }  ,fill opacity=0.64 ]   (170.65,49.79) -- (276.71,155.85) ;
		\draw [shift={(170.65,49.79)}, rotate = 45] [color={rgb, 255:red, 0; green, 0; blue, 0 }  ][line width=0.75]      (0, 0) circle [x radius= 3.35, y radius= 3.35]   ;
		\draw [fill={rgb, 255:red, 155; green, 155; blue, 155 }  ,fill opacity=0.64 ]   (170.65,49.79) -- (170.65,261.91) ;
		\draw [fill={rgb, 255:red, 155; green, 155; blue, 155 }  ,fill opacity=0.64 ]   (170.65,261.91) -- (276.71,155.85) ;

		\draw (148,142.4) node [anchor=north west][inner sep=0.75pt]    {$\Gamma $};
		\draw (149,31.4) node [anchor=north west][inner sep=0.75pt]    {$\mathcal{O}$};

		
		\coordinate (R1) at (244.77,187.79);
		\coordinate (R2) at (265.98,166.58);
		
\coordinate (R3) at (203.98,104.58);
\coordinate (R4) at (182.77,125.79);
		
		\draw[thick] (R1)--(R2)--(R3)--(R4)--cycle;
		
		\filldraw (R3) circle (2pt);
		\node[left] at (R3) {$P$};
		
	
	\draw[<->] 
	(241.77,184.79) -- (262.98,163.58);
	
	\node[rotate=45] at (245.38,167.00) {$\delta$};
	
		\node at (215.38,137.00) {$C_{P,\delta}$};
		
	\end{tikzpicture}
\caption{The region $C_{P,\delta}$ of width $\delta$.}
\end{figure}
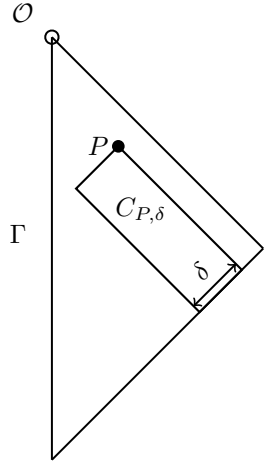

For any $(u,v)\in C_{P,\delta}$,
\begin{equation*}
    |r(u,v)-r(u,v_0)| \lesssim |v-v_0| \leq \delta \leq \delta' F(P)^{-1}r_0 \leq \delta' r_0 \leq \delta' r(u,v_0),
\end{equation*}  
and therefore $r(u,v)/r(u,v_0)\sim1$ uniformly when $\delta'$ is sufficiently small.

We first estimate the metric and scalar-field quantities satisfying $v$-transport equations.
\begin{align*}
    |\log(-\nu)(u,v) - \log(-\nu)(u,v_0)| \leq  \int_{(u,v)}^{(u,v_0)}\frac{\mu\lambda}{\rho r}dv \lesssim \delta r(u,v_0)^{-1},
\end{align*}
\begin{align*}
    |m(u,v)-m(u,v_0)| &\lesssim \int_{(u,v)}^{(u,v_0)} \frac{\rho}{2\lambda}(r\partial_v\phi)^2 dv \\ 
    &\lesssim C^2 F(P)^2 \delta.
\end{align*}
\begin{align*}
   | \log(\rho^{-1}(-\nu))(u,v_0)- \log(\rho^{-1}(-\nu))(u,v) | &\leq \int_{(u,v)}^{(u,v_0)} \frac{1}{\lambda}r(\partial_v\phi)^2 dv \\
   &\lesssim \int_{(u,v)}^{(u,v_0)} C^2 F(P)^2 r^{-1} dv\\
   &\lesssim  \delta C^2 F(P)^2  r(u,v_0)^{-1} . 
\end{align*}
\begin{align*}
   | (r\partial_u\phi)(u,v_0)-(r\partial_u\phi)(u,v) | &\leq \int_{(u,v)}^{(u,v_0)} (-\nu) |\partial_v\phi| dv \\
   &\lesssim (-\nu)(u,v_0)  C F(P)\int_{(u,v)}^{(u,v_0)}r^{-1} dv\\
   &\lesssim  \delta (-\nu)(u,v_0)  C F(P) r(u,v_0)^{-1}. 
\end{align*}

In particular,
\begin{align*}
    |\partial_u\phi(u,v_0)-\partial_u\phi(u,v)| &\leq \frac{1}{r(u,v)}| (r\partial_u\phi)(u,v_0)-(r\partial_u\phi)(u,v) | + |r(u,v_0)\partial_u\phi(u,v_0)||r(u,v)^{-1}-r(u,v_0)^{-1}| \\
    &\lesssim \delta (-\nu)(u,v_0)  C F(P) r(u,v_0)^{-2} + \delta  r(u,v_0)^{-2} | r(u,v_0) \partial_u\phi(u,v_0)| \\
    &\lesssim \delta' (-\nu)(u,v_0) C  r(u,v_0)^{-1} + \delta' F(P)^{-1}|\partial_u\phi(u,v_0)|
\end{align*}

We next consider the quantities $\lambda, r\partial_v\phi$, which satisfy transport equations in $u$. To compare $\lambda$ with $\lambda(u,v_0)$, we integrate along the ingoing cones $\underline{\Sigma}_{v_0}$ and $\underline{\Sigma}_v$ and compare the resulting expressions:
\begin{align*}
    \log\lambda(u,v_0)-\log\lambda(u,v) &= \log\lambda(-1,v_0)-\log\lambda(-1,v) + \int_{-1}^u \Big(\frac{2m \nu}{\rho r^2}(u',v_0)- \frac{2m \nu}{\rho r^2}(u',v)\Big)du' \\
    &= O(|v-v_0|) + \int_{-1}^u 2\big((\rho^{-1}\nu)r^{-2}\big)(u',v_0)\Big(m(u',v_0)- m(u',v)\Big)du' \\
    & +\int_{-1}^u 2 m(u',v) r^{-2}(u',v_0)\Big((\rho^{-1}\nu)(u',v_0)- (\rho^{-1}\nu)(u',v)\Big)du' \\
    & +\int_{-1}^u 2 m(u',v)(\rho^{-1}\nu)(u',v)\Big((r^{-2})(u',v_0)- (r^{-2})(u',v)\Big)du'.
\end{align*}
Inserting the difference estimates gives 
\begin{align*}
|\log\lambda(u,v_0)-\log\lambda(u,v)| &\lesssim \delta + \int_{-1}^{u} \delta (-\nu)(u,v_0) C^2 F(P)^2 r(u,v_0)^{-2} du \\
&\lesssim  \delta + \delta  C^2 F(P)^2 r(u,v_0)^{-1} \\
&\lesssim \delta + C^2 \delta' F(P).
\end{align*}

We apply the same comparison argument to $r\partial_v\phi$:
\begin{align*}
     (r\partial_v\phi)(u,v_0)-r\partial_v\phi(u,v) &= (r\partial_v\phi)(-1,v_0)-r\partial_v\phi(-1,v) + \int_{-1}^{u}\Big((-\lambda)\partial_u\phi(u',v_0)- (-\lambda)\partial_u\phi(u',v) \Big)du' \\
     &= O(|v-v_0|) + \int_{-1}^{u}\partial_u\phi(u',v_0) \Big((-\lambda)(u',v_0)-(-\lambda)(u',v) \Big)  du \\
     &\hspace{5em}+ \int_{-1}^{u}(-\lambda)(u',v) \Big(\partial_u\phi(u',v_0) -\partial_u\phi(u',v)\Big)  du,
\end{align*}
and so, with all integrands below evaluated at $(u',v_0)$,
\begin{align*}
   |(r\partial_v\phi)(u,v_0)-r\partial_v\phi(u,v)| &\lesssim \delta + \delta C^2 F(P)^2\int_{-1}^u r^{-1}|\partial_u\phi|\,du' \\
   &\qquad + \delta C F(P)\int_{-1}^u\frac{-\nu}{r^2}\,du' \\
   &\lesssim \delta + \delta C F(P)r_0^{-1} \\
   &\qquad + \delta C^2 F(P)^2\Big(\int_{-1}^u\frac{-\nu}{r^3}\,du'\Big)^{\frac12}\Big(\int_{-1}^u\frac{r(\partial_u\phi)^2}{-\nu}\,du'\Big)^{\frac12} \\
   &\lesssim \delta +\delta C F(P)r_0^{-1} + \delta C^2  F(P)^2 r_0^{-1}  \\
   &\lesssim \delta + C \delta' + C^2 F(P)\delta' .
\end{align*}
Here \eqref{eq:EP:SSESF15} and \eqref{eq:3.1}--\eqref{eq:3.2} give
\[
    \int_{-1}^u\frac{r(\partial_u\phi)^2}{-\nu}\,du'
    =\Big[\log\frac{\rho}{\lambda}\Big]_{-1}^u\lesssim1,
\]
while $\partial_u r=\nu$ yields
\[
    \int_{-1}^u\frac{-\nu}{r^3}\,du'
    =\Big[\frac{1}{2r^2}\Big]_{-1}^u\leq\frac{1}{2r_0^2}.
\]
Choosing $\delta'$ sufficiently small in terms of $C$ improves the bootstrap bound and closes the argument. Estimate \eqref{eq:stabilityest} follows.
\end{proof}

We now use the local stability estimate to obtain global pointwise bounds.

\begin{proposition}\label{prop:pointwise_bounds}
    The following estimates hold:
    \begin{align}
        & \sup_{\mathcal{Q}}(|\phi| + |rD_v\phi| + |rD_u\phi| + |r^2D_{vv}^2\phi|) \leq C, \\
        & |rD_v\phi| \lesssim \mu^{\frac13}.
    \end{align}
\end{proposition}
\begin{proof}
To see that $rD_v\phi$ is globally bounded, assume the contrary, namely that there exists a sequence $p_i=(u_i,v_i)$ with $u_i\nearrow0$ such that $c_i\doteq|rD_v\phi(u_i,v_i)|\to\infty$. Without loss of generality, we may assume that $1+c_i=F(p_i)$, where $F$ is defined in Lemma \ref{lem:local}.
    
    Let $r_i\doteq r(u_i,v_i)$. Lemma \ref{lem:local} gives a sufficiently small $\delta'>0$, independent of $i$, such that
    \begin{equation*}
        |r\partial_v\phi(u_i,v)-r\partial_v\phi(u_i,v_i)| \leq \frac12 c_i
    \end{equation*}
    holds for $v \in I_i \doteq [v_i - \delta' c_i^{-1}r_i, v_i]$. In particular, we have the lower bound $|r\partial_v\phi(u,v)| \geq \frac12 c_i$ in $I_i$. Computing the Hawking mass in this interval gives 
    \begin{align*}
        m(u_i,v_i) &\geq m(u_i,v_i)-m(u_i,v_i-\delta' c_i^{-1}r_i) \\
        &\gtrsim \int_{v_i-\delta' c_i^{-1}r_i}^{v_i} (r\partial_v\phi)^2(u_i,v')dv'\\
        &\gtrsim \delta' c_i r_i.
    \end{align*}
    This would imply 
    \begin{align*}
        \mu(u_i,v_i) = \frac{2m(u_i,v_i)}{r_i} \gtrsim \delta' c_i \rightarrow \infty,
    \end{align*}
a contradiction.

\vspace{1em}
\noindent
Therefore $r\partial_v\phi$ is uniformly bounded. Next, the identity
\begin{align*}
    \phi(u,v)-\phi(-1,v) &= \int_{-1}^{u} \partial_u\phi(u',v) du' \\
    &= \int_{-1}^{u} \frac{(-\nu)\mu}{\rho r} (rD_v\phi)(u',v)du' + \big( (rD_v\phi)(-1,v)-(rD_v\phi)(u,v) \big),
\end{align*}
and the bound on the Doppler exponent show that the right-hand side is uniformly bounded. Hence $\phi$ is uniformly bounded.

\vspace{1em}
\noindent
To control the incoming derivative, we integrate the wave equation
\begin{align*}
    \partial_v \partial_u (r\phi) = \frac{\mu \lambda \nu}{\rho r}\phi,
\end{align*}
from the axis, where $\partial_u(r\phi)=\nu\phi=-\lambda\phi$ is uniformly bounded. This gives
\begin{align*}
    \partial_u(r\phi)(u,v)& = -\lambda \phi(u,u) + \int_{u}^{v} \frac{\mu \lambda \nu}{\rho r}\phi(u,v')dv',
\end{align*}
which can be estimated as 
\begin{align*}
    |\partial_u (r\phi)(u,v)| &\lesssim \lambda |\phi(u,u)| + \sup_{\Sigma_u, v' \leq v}| \phi(u,v')|\int_{u}^{v}\frac{\mu \lambda (-\nu)}{\rho r}dv\\
    &\lesssim 1 + (-\nu)(u,v)-(-\nu)(u,u) \\
    &\lesssim 1 + (-\nu)(u,v).
\end{align*} 
It follows that 
\begin{align*}
    |r\partial_u\phi(u,v) + \nu\phi(u,v)| \lesssim 1 + (-\nu)(u,v),
\end{align*}
and so $|rD_u\phi| \lesssim 1$.

We next refine the bound for $rD_v\phi$ in terms of the local value of $\mu$. Let $P=(u,v)$ and set $|r\partial_v\phi|(u,v)=\kappa\geq0$. By Lemma \ref{lem:local}, there exists a uniform constant $\delta_0>0$ such that, whenever $\delta'\leq\delta_0$, the interval $I=[v-\delta'F(P)^{-1}r(u,v),v]$ satisfies
\begin{align} \label{discuss kappa}
|r\partial_v \phi|(u,v') &\geq \kappa-C\delta' F(P)\\ \nonumber
&\geq \kappa-\frac14 F(P),
\end{align}
where $C$ is the uniform constant in \eqref{eq:stabilityest}. Since $rD_v\phi$ is already bounded, $F(P)$ is bounded by a fixed large constant $F$. We consider two cases. If $\kappa\geq\frac12F$, take $\delta'=\delta_0$. Then \eqref{discuss kappa} gives $|r\partial_v\phi|\geq\frac12\kappa$ throughout the interval, and hence
\begin{align*}
	m(u,v) &\gtrsim \int_{v-\delta' F(P)^{-1} r(u,v)}^{v}(r\partial_v\phi)^2\\
	&\gtrsim \delta'F(P)^{-1} r(u,v)\kappa^2  \\
	&\geq \frac12 \delta_0 \kappa r(u,v).
\end{align*}
It follows that $\mu^\frac13\gtrsim\mu \gtrsim \kappa = |r\partial_v\phi|$, as desired.

If $\kappa<\frac12F$, choose $\delta'=\min\{\delta_0,\frac12(CF)^{-1}\kappa\}$. Then $|r\partial_v\phi|\geq\frac12\kappa$ throughout the interval. If $\delta'=\delta_0$, then
\begin{align*}
	m(u,v) &\gtrsim \int_{v-\delta' F(P)^{-1} r(u,v)}^{v}(r\partial_v\phi)^2\\
	&\gtrsim \delta_0F^{-1} r(u,v)\kappa^2 .
\end{align*}
It follows that $\mu^\frac13\gtrsim\mu^\frac12 \gtrsim \kappa = |r\partial_v\phi|$, as desired.

If $\delta'=\frac12(CF)^{-1}\kappa$, it follows that 
\begin{align*}
	m(u,v) &\gtrsim \int_{v-\delta' F(P)^{-1} r(u,v)}^{v}(r\partial_v\phi)^2\\
	&\gtrsim \delta'F(P)^{-1} r(u,v)\kappa^2  \\
	&\gtrsim F^{-2} \kappa^3 r(u,v).
\end{align*}
It follows that $\mu \gtrsim \kappa^3 = |r\partial_v\phi|^3$, as desired.

\vspace{1em}
\noindent
Finally, we prove the higher-order bound for $r^2D_{vv}^2\phi$. In view of the bound for $rD_v\phi$ and \eqref{alpha' relation}, it suffices to show $|r\alpha'|\lesssim1$, where $\alpha'=D_{vv}^2(r\phi)$.

We compute 
\begin{align*}
    \partial_u(\lambda^2  \alpha') =  \frac{(-\nu)}{\lambda \rho} r (\partial_v\phi)^3 + \frac{3\nu \mu \lambda}{\rho r} (\partial_v\phi).
\end{align*}
Integrating this expression from the initial data gives 
\begin{align*}
    |\lambda^2 r\alpha'(u,v)| &\lesssim |\lambda^2 \alpha'(-1,v)|\cdot r(u,v) + \int_{-1}^{u} \Big( r^2 |\partial_v\phi|^3 + \mu |\partial_v\phi|  \Big) (-\nu) du' \\
    &\lesssim r(u,v)+ \int_{-1}^{u}\Big( \frac{\mu}{r} + \frac{\mu^{\frac43}}{r}\Big)(-\nu)du' \\
    &\lesssim 1,
\end{align*}
by the boundedness of the Doppler integral. The result follows.

\end{proof}
The preceding pointwise estimates also yield decay along the final cone $\underline{\Sigma}_0$.
\begin{proposition}
	Along $\{v=0\}$, we have
	\begin{align}
		&\lim_{u\nearrow0}\mu(u,0)=0, \label{eq:ext_prin_mutozero_selfsim}\\
		&\lim_{u\nearrow0}(rD_v\phi)(u,0)=0. \label{eq:3.6}
	\end{align}
	Here and below, a subscript $0$ denotes restriction to $\underline{\Sigma}_0$.
\end{proposition}
\begin{proof}
	The first statement was proved by Christodoulou in \cite{christ4}. We outline the argument in the present notation. By \eqref{eq:finite_blueshift}, passing to $v=0$ on compact $u$-intervals and then letting the upper endpoint tend to $0$, we obtain
	\[
		\int_{-1}^{0}\frac{\mu_0(-\nu_0)}{\rho_0r_0}\,du<\infty.
	\]
	For $-1\leq u_2<u<0$, set
	\[
		\gamma_0(u;u_2):=\int_{u_2}^{u}\frac{\mu_0(-\nu_0)}{\rho_0r_0}\,du'.
	\]
	Since $\partial_u m\leq0$ by \eqref{eq:SSESF2}, we have
		\begin{equation}\label{FIT58}
			\begin{aligned}
				\gamma_0(u;u_2)
				&\geq \int_{r_0(u)}^{r_0(u_2)} \frac{\mu_0(u)r_0(u)}{r_0\cdot(r_0-\mu_0(u)r_0(u))}\;\D r_0 \\
				&= \int_{r_0(u)}^{r_0(u_2)} \frac{1}{r_0-\mu_0(u)r_0(u)}-\frac{1}{r_0}\;\D r_0 \\
				&= \log \left(\frac{r_0(u)}{r_0(u_2)}+\frac{r_0(u_2)-r_0(u)}{r_0(u_2)-\mu_0(u)r_0(u_2)}\right).
			\end{aligned}
		\end{equation}
		The above estimate implies that
		\begin{equation*}
			\frac{1}{1-\mu_0(u)}\leq \frac{1}{r_0(u_2)-r_0(u)}\ml r_0(u_2)e^{\gamma_0(u;u_2)}- r_0(u) \mr,
		\end{equation*}
		which further yields
		\begin{equation}\label{FIT222}
			1-\mu_0(u)\geq e^{-\gamma_0(u;u_2)}+\frac{e^{-\gamma_0(u;u_2)}-1}{e^{\gamma_0(u;u_2)}\frac{r_0(u_2)}{r_0(u)}-1}.
		\end{equation}
		Since $r_0$ is continuous and strictly decreasing to $0$, for each $u_2$ we may choose a unique $u=u(u_2)>u_2$ such that $r_0(u)/r_0(u_2)=\frac12$. As $u_2\nearrow0$, we have $u(u_2)\nearrow0$ and $\gamma_0(u(u_2);u_2)\to0$ by the integrability above. The right-hand side of \eqref{FIT222} therefore tends to $1$, so $\mu_0(u(u_2))\to0$. Since $u(u_2)$ ranges over a full interval ending at $0$, this proves $\lim_{u\nearrow0}\mu_0(u)=0$.

	To prove the second statement, suppose instead that there is an increasing sequence $u_i \rightarrow 0$ and a constant $c_0 > 0$ such that $r\partial_v\phi(u_i,0) \geq c_0 > 0$. 
	
	Apply \eqref{eq:stabilityest} in Lemma \ref{lem:local} to compare $r\partial_v\phi(u,v)$ with its value on $\{v=0\}$ whenever $|v| \lesssim \epsilon r_0(u)\lesssim \epsilon |u|$, with $\epsilon>0$ sufficiently small. Since $r\partial_v\phi(u,0)$ is uniformly bounded, this gives 
	\begin{equation*}
		|r\partial_v\phi(u,v)-r\partial_v\phi(u,0)| \lesssim \epsilon.
	\end{equation*}
	Setting $u = u_i$ and choosing $\epsilon$ sufficiently small, we obtain
	\begin{equation*}
		|r\partial_v\phi(u_i,v)| \geq \frac12 c_0.
	\end{equation*}
	The Hawking mass therefore satisfies
	\begin{align*}
		m(u_i,0) &\gtrsim \int_{-\epsilon F(u_i,0)^{-1} r_0(u)}^{0}(r\partial_v\phi)^2\\
		&\gtrsim \epsilon c_0^2  r_0(u),
	\end{align*} 
	and hence, for every $i$,
	$$\mu(u_i,0) \gtrsim \epsilon c_0^2.$$ 
	However, this contradicts (\ref{eq:ext_prin_mutozero_selfsim}).
\end{proof}
\section{Self-similar estimates} \label{sec:selfsimilar}
In this section, we obtain refined estimates at the self-similar scale. Following Christodoulou \cite{christ3}, we introduce the self-similar variable
\[
    x\triangleq\frac{r}{-u}.
\]
For fixed $0<\delta\ll1$, define the exterior and interior self-similar regions by
\[
    \mathcal C_\delta=\mathcal Q\cap\{x\geq\delta\},
    \qquad
    \mathcal B_\delta=\mathcal Q\cap\{0\leq x\leq\delta\},
\]
and let $\Gamma_\delta=\{x=\delta\}$ denote their common boundary. Choose $\delta<\min\{c,c_1\}$, where $c,c_1>0$ are the lower-bound constants in \eqref{eq:3.3}--\eqref{eq:3.4}. Since $\lambda>0$, this boundary has a unique parametrization
\[
    r(u,v_\delta(u))=\delta|u|,\qquad -1\leq u<0,
\]
with $u<v_\delta(u)<0$. Differentiation gives $v_\delta'(u)=(-\nu-\delta)/\lambda>0$, so $\Gamma_\delta$ is timelike. We write $u_\delta$ for the inverse of $v_\delta$ on its range, with the limiting convention $u_\delta(0)=0$. In particular,
\[
    \mathcal C_\delta=\{(u,v):-1\leq u<0,\ v_\delta(u)\leq v\leq0\}.
\]

\begin{figure}[htbp]
\centering
\tikzset{every picture/.style={line width=0.75pt}} 

\begin{tikzpicture}[x=0.75pt,y=0.75pt,yscale=-1,xscale=1]
	
	\draw [fill={rgb, 255:red, 155; green, 155; blue, 155 }  ,fill opacity=0.64 ]   (172.31,51.45) -- (276.71,155.85) ;
	\draw [shift={(170.65,49.79)}, rotate = 45] [color={rgb, 255:red, 0; green, 0; blue, 0 }  ][line width=0.75]      (0, 0) circle [x radius= 3.35, y radius= 3.35]   ;
	\draw [fill={rgb, 255:red, 155; green, 155; blue, 155 }  ,fill opacity=0.64 ]   (170.65,49.79) -- (170.65,261.91) ;
	\draw [fill={rgb, 255:red, 155; green, 155; blue, 155 }  ,fill opacity=0.64 ]   (170.65,261.91) -- (276.71,155.85) ;
	\draw    (170.65,49.79) -- (208,224) ;
	
	\draw (148,142.4) node [anchor=north west][inner sep=0.75pt]    {$\Gamma $};
	\draw (149,31.4) node [anchor=north west][inner sep=0.75pt]    {$\mathcal{O}$};
	\draw (212,129.4) node [anchor=north west][inner sep=0.75pt]    {$\mathcal{C}_{\delta }$};
	\draw (183,89.4) node [anchor=north west][inner sep=0.75pt]    {$\Gamma _{\delta }$};
	\draw (173,164.4) node [anchor=north west][inner sep=0.75pt]    {$\mathcal{B}_{\delta }$};

\end{tikzpicture}
	\caption{Interior and exterior self-similar regions.}
\label{FETFig2}
\end{figure}
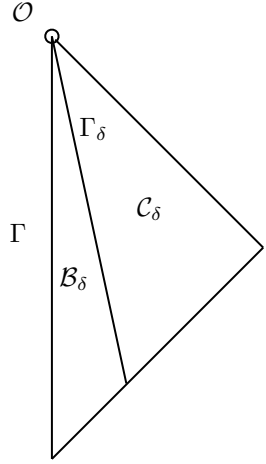
We first propagate the estimates from $\underline\Sigma_0$ into the exterior self-similar region.
\begin{lemma}
	For any fixed $\delta \in (0,1)$, we have
	\begin{align}
		r(u,v) \sim_\delta |u| \ \text{in} \  \mathcal{C}_{\delta}, \label{eq:3.8}
	\end{align}
	as well as 
	\begin{align}
		\lim_{u \rightarrow 0}\sup_{\mathcal{C}_{\delta}\cap\{u' \geq u\}}\mu(u',v) &= 0, \label{eq:3.9}\\
		\sup_{\mathcal{C}_{\delta}} |r\partial_v\phi| &\lesssim_\delta 1, \label{eq:3.10}\\
		\sup_{\mathcal{C}_{\delta}} |\phi| &\lesssim_\delta 1. \label{eq:3.11}
	\end{align}
\end{lemma}
\begin{proof}
	Since $\lambda$ is bounded above and below by positive constants, we have 
	\begin{align*}
		r(u,v) \sim v-u \sim |u|\big(1-\frac{v}{u}\big).
	\end{align*}
	Moreover, $\big(1-\frac{v}{u}\big) \sim_\delta 1$ in $\mathcal{C}_\delta$, which gives \eqref{eq:3.8}.

	To prove \eqref{eq:3.9}, observe that, for every $(u,v)\in\mathcal C_\delta$,
	\begin{align*}
		\mu(u,v) = \frac{2 m(u,v)}{r(u,v)} \leq \Big(\frac{r(u,0)}{r(u,v)} \Big)\frac{2 m(u,0)}{r(u,0)}.
	\end{align*}
	By definition, $(u,v) \in \mathcal{C}_\delta$ implies $r(u,v) \geq \delta |u|$. Moreover, $r(u,0) \sim |u|$. Hence
	\begin{align*}
		\Big(\frac{r(u,0)}{r(u,v)} \Big)\frac{2 m(u,0)}{r(u,0)} \lesssim_\delta \mu(u,0),
	\end{align*}
	with the latter tending to zero by \eqref{eq:ext_prin_mutozero_selfsim}.
	
	We next prove \eqref{eq:3.11}. The fundamental theorem of calculus and Cauchy--Schwarz give
	\begin{align*}
		|\phi(u,v) - \phi_0(u)| &\leq \int_{v}^{0}|\partial_v\phi(u,v')|dv' \\
		&\lesssim \Big(\int_{v}^{0} r(u,v')^{-2}dv' \Big)^{\frac12}\Big(\int_{v}^{0} (r \partial_v\phi)^2(u,v')dv' \Big)^{\frac12} \\
		&\lesssim r(u,v)^{-\frac12} m_0(u)^{\frac12} \\
		&\lesssim_\delta \mu_0(u)^{\frac12}.
	\end{align*}
	
	Finally, integrate \eqref{eq:SSESF11} in the ingoing direction and use \emph{(F)} and \eqref{eq:3.11} to obtain
	\begin{align*}
		|\partial_v(r\phi)(u,v)-\partial_v(r\phi)(-1,v)| &\lesssim \int_{-1}^{u} \frac{\mu(-\nu)}{\rho r}(u',v) |\phi|(u',v)du'\\
		&\lesssim_\delta 1.
	\end{align*}
	Thus $\partial_v(r\phi)=\lambda\phi+r\partial_v\phi$ is bounded pointwise in $\mathcal C_\delta$. Since the first term is already controlled, \eqref{eq:3.10} follows.
\end{proof}

The same monotonicity properties yield the following integrated estimates.
\begin{proposition} \label{ext bulk}
	For any fixed $\delta>0$ sufficiently small as specified above, we have 
	\begin{align}
		\label{eq:3.12}
		\int_{\Gamma_\delta} \frac{(-\nu) \mu}{\rho r} du + \iint_{\mathcal{C}_{\delta}} \Big(\frac{m}{r^3} + (\partial_v\phi)^2\Big) (-\nu)dudv  \lesssim_\delta 1.
	\end{align}
\end{proposition}
\begin{proof}
	For $(u,v)\in\mathcal C_\delta$, we have $r\sim_\delta r_0\sim_\delta|u|$ and, by monotonicity, $m(u,v)\leq m_0(u)$ and $(-\nu)(u,v)\leq(-\nu_0)(u)$. Therefore,
	\begin{align}
		\iint_{\mathcal{C}_\delta}\frac{m}{r^3}(-\nu)dudv &= \int_{-1}^{0}\int_{v_\delta(u)}^{0} \frac{m}{r^3}(-\nu)dv du \nonumber \\
		&\lesssim_\delta \int_{-1}^{0}\frac{m_0}{|u|^{3}}(-\nu_0)\Big(\int_{v_\delta(u)}^{0}dv \Big)du \nonumber \\
		&\lesssim_\delta \int_{-1}^{0} \frac{m_0}{|u|^{2}}(-\nu_0)du \nonumber \\
		&\lesssim_\delta \int_{-1}^{0} \frac{(-\nu_0)\mu_0}{r_0}du \nonumber \\
		&\lesssim_\delta 1, \label{eq:temp2}
	\end{align}
	using $-v_\delta(u)\leq|u|$, $r_0\sim|u|$, and (F).
	
	Next, we use the identity
	\begin{align}
		\label{temp2.5}
		\partial_v\Big(\frac{(-\nu)\mu}{\rho r} \Big) = \frac{(-\nu)}{\lambda \rho}(\partial_v\phi)^2  - \frac{2 \lambda (-\nu) \mu}{\rho r^2}.
	\end{align}
	Integrating this identity in $\mathcal{C}_\delta$ gives 
	\begin{align*}
		\int_{\underline{\Sigma}_0}\Big(\frac{(-\nu)\mu}{\rho r} \Big) du - \int_{\Gamma_\delta}\Big(\frac{(-\nu)\mu}{\rho r} \Big) du = \iint_{\mathcal{C}_\delta} \frac{(-\nu)}{\lambda \rho}(\partial_v\phi)^2 dudv-\iint_{\mathcal{C}_\delta} \frac{2 \lambda (-\nu) \mu}{\rho r^2} dudv,
	\end{align*}
	or equivalently, using (F) and (\ref{eq:temp2}),
	\begin{align*}
		\int_{\Gamma_\delta}\Big(\frac{(-\nu)\mu}{\rho r} \Big) du + \iint_{\mathcal{C}_\delta} \frac{(-\nu)}{\lambda \rho}(\partial_v\phi)^2 dudv &\leq \int_{\underline{\Sigma}_0}\Big(\frac{(-\nu)\mu}{\rho r} \Big) du + \iint_{\mathcal{C}_\delta} \frac{2 \lambda (-\nu) \mu}{\rho r^2} dudv \\
		&\lesssim 1 + \iint_{\mathcal{C}_\delta} \frac{m}{r^3} (-\nu) dudv \\
		&\lesssim 1.
	\end{align*}
	Since $\lambda$ and $\rho$ are bounded above and below, this proves the remaining bounds in \eqref{eq:3.12}.
\end{proof}
We next refine the estimates near $\mathcal O$. Define the tail of the Doppler integral along the final cone by
\[
    \gamma_0(u):=\int_u^0\frac{\mu_0(-\nu_0)}{\rho_0r_0}(u')\,du'.
\]
The integrability along this cone established in Section \ref{subsec:extprin_selfsimilarregion} implies $\gamma_0(u)\to0$ as $u\nearrow0$. Thus, by \eqref{eq:3.9}, for every sufficiently small $\epsilon>0$ we may choose $u_0^{(\epsilon)}<0$ such that $\gamma_0(u_0^{(\epsilon)})\leq\epsilon$ and $\mu(u,v)\leq\epsilon$ throughout
	$$
	\mathcal{C}^{(\epsilon)}_\delta=\mathcal{C}_\delta \cap \{u\geq u_0^{(\epsilon)}\}.
	$$
	\begin{remark}
		The choice of $u_0^{(\epsilon)}$ may depend on $\delta$. Since the tail is nonincreasing, $\gamma_0(u)\leq\epsilon$ for all $u\geq u_0^{(\epsilon)}$.
	\end{remark}
	In this smaller region, we have the following preliminary estimates.
	\begin{lemma}
		Define
		$$
		\tilde{\gamma}(u;v,0):=\int_v^0 \frac{\mu\lambda}{r\rho}(u,v') dv', \quad \gamma(v;a,b):=\int_a^b \frac{\mu(-\nu)}{r\rho}(u',v) du'.
		$$
		Here $-1\leq a\leq b<v\leq0$; on $v=0$, the endpoint $b=0$ is also allowed, with the integral understood in the improper sense.
		The following estimates hold in $\mathcal{C}^{(\epsilon)}_\delta$:
		\begin{align}
			&\tilde{\gamma}(u;v,0)\lesssim_\delta \epsilon, \label{tgamma Sdelta}\\
			&\gamma(v;u_0^{(\epsilon)},u_\delta(v))\lesssim_\delta \epsilon, \label{gamma Sdelta}\\
			&-\nu(u,v)\approx -\nu_0(u), \label{nu Sdelta}\\
			&\lambda(u,v) \approx \lambda(u_0^{(\epsilon)},v),\label{lambda Sdelta}\\
			&\int_{u_0^{(\epsilon)}}^u r\frac{\rho (\partial_u \phi)^2}{-\nu}(u',v) du'\lesssim_\delta \epsilon, \quad  \int_{v_\delta(u)}^0 r\frac{\rho (\partial_v \phi)^2}{\lambda}(u,v') dv'\lesssim_\delta \epsilon. \label{cor} 
		\end{align}
	\end{lemma}
	\begin{proof}
		First,
		$$
		\tilde{\gamma}(u;v,0)\leq \epsilon \int_v^0 \frac{1}{r\rho}(u,v') dv'\lesssim_\delta \epsilon.
		$$
		Applying \eqref{tgamma Sdelta} to \eqref{eq:SSESF1} and using monotonicity, we obtain
		$$
		-\nu(u,v)\leq -\nu_0(u) \lesssim -\nu(u,v)e^{C_\delta \epsilon}.
		$$
		Consequently,
		$$
		\gamma(v;u_0^{(\epsilon)},u_\delta(v))
            \lesssim_\delta \int_{u_0^{(\epsilon)}}^{u_\delta(v)}
            \frac{\mu_0(-\nu_0)}{\rho_0r_0}(u')\,du'
            \leq \gamma_0(u_0^{(\epsilon)})\leq\epsilon,
		$$
		which proves \eqref{gamma Sdelta}. The same argument gives \eqref{lambda Sdelta}. Finally, integrate \eqref{eq:SSESF4}--\eqref{eq:SSESF5} and use $\mu\leq\epsilon$ together with \eqref{tgamma Sdelta}--\eqref{gamma Sdelta} to obtain \eqref{cor}. 
	\end{proof}
	\begin{remark}
		The first estimate in \eqref{nu Sdelta} also gives smallness of $|\frac{1}{\nu_0}-\frac{1}{\nu}|$. It does not, however, control $|\nu_0-\nu|$, since we do not yet have an upper bound for $-\nu$.
	\end{remark}
	\begin{remark}
		For every fixed $\delta>0$ sufficiently small as specified above, $\lambda$ extends continuously to $\mathcal O$ within $\mathcal C_\delta$, with
		\begin{equation}\label{lambda conti}
			\lim_{\substack{(u,v)\to(0,0)\\(u,v)\in\mathcal C_\delta}}\lambda(u,v)
			=\lambda_*:=\lim_{u\to 0}\lambda_0(u)>0.
		\end{equation}
		Indeed, the limit along the final cone exists and is positive by monotonicity and Lemma \ref{lem:extprin_1}. For $a\leq u$ and $(u,v)\in\mathcal C_\delta$, the same comparison with the terminal cone as above gives
		\[
			|\log\lambda(u,v)-\log\lambda(a,v)|
			=\gamma(v;a,u)\lesssim_\delta\gamma_0(a).
		\]
		Continuity along $u=a$, followed by $a\to 0$, proves \eqref{lambda conti}. This observation is not used in the subsequent arguments.
	\end{remark}

The estimates in this section exclude singularity formation at the self-similar scale. Any remaining obstruction to extension must therefore arise through concentration of $\mu$ or $|rD_v\phi|$ at a scale strictly faster than the self-similar scale. The following sections rule out precisely this scenario.

\section{Higher-order energy identities}
\label{subsec:extprin_multipliers}
In this section, we derive a family of higher-order weighted energy identities. Their coercivity determines the thresholds $\frac14$ and $\frac38$.

We begin with the general identity.
\begin{proposition} \label{prop:energy}
    Let $\tilde\Gamma(u,v)$ be sufficiently regular and set $\Theta=r\lambda\tilde\Gamma(\alpha')^2$. Then
    \begin{align}\label{eq:ext_prin_mainenergyestimate}
    \partial_u \Theta + G_1(D_{vv}^2\phi)^2 + G_2(D_v\phi)^6 + G_3(D_v\phi)^4 + G_4(D_v\phi)^2 = \partial_v B ,
\end{align}
where the coefficients $G_i$ are given by
\begin{align*}
    G_1 &= \Big(-r\partial_u \tilde{\Gamma} + \frac{(-\nu)(1-4\mu)}{\rho}\tilde{\Gamma}\Big)r^2 \lambda, \\
    G_2 &= \frac{r^4 (-\nu)\lambda}{2 \rho}\tilde{\Gamma}, \\
    G_3 &= \frac{r^2 (-\nu)\lambda}{2 \rho}\tilde{\Gamma} + \frac{r^3(-\nu)}{2 \rho}\partial_v \tilde{\Gamma},  \\
    G_4 &= \frac{2(-\nu)\lambda(1-2\mu)}{\rho}\tilde{\Gamma} - \frac{r (-\nu)(2-5\mu)}{ \rho}\partial_v \tilde{\Gamma} + 2 r^2 \partial_u\partial_v\tilde{\Gamma}.
\end{align*}
and the boundary flux $B$ is
\begin{align*}
	B &= -\frac{r^3 \nu}{2 \rho} \tilde{\Gamma} (D_v\phi)^4 - \frac{5 r \nu \mu}{ \rho}\tilde{\Gamma} (D_v\phi)^2 + \frac{2r\nu}{\rho}\tilde{\Gamma}(D_v\phi)^2+ 2r^2 \partial_u \tilde{\Gamma} (D_v\phi)^2 .
\end{align*}
\end{proposition}
\begin{proof}
	Using \eqref{alpha' relation} and \eqref{eq:SSESF18}, we compute
	\begin{align*}
		\partial_u\Theta &= \nu \lambda \tilde{\Gamma} \alpha'^2 + r \big(\frac{\mu\lambda\nu}{\rho r}\big)\tilde{\Gamma} \alpha'^2 + r \lambda \partial_u \tilde{\Gamma} \alpha'^2 + 2 r \lambda \tilde{\Gamma} \alpha' \partial_u \alpha' \\
		&= \big(\nu \lambda \tilde{\Gamma} + \frac{\mu}{\rho}\lambda\nu \tilde{\Gamma} + r \lambda \partial_u \tilde{\Gamma} \big)(rD_{vv}^2\phi + 2 D_v\phi)^2 \\
		&\quad + 2 r \lambda \tilde{\Gamma} (rD_{vv}^2\phi + 2 D_v\phi)\big(\frac{2\mu(-\nu)}{\rho} D_{vv}^2\phi + \frac{(-\nu)}{\rho}r (D_v\phi)^3 + \frac{(-\nu)\mu}{\rho r}D_v\phi \big).
	\end{align*}
	Expanding and reorganizing the right-hand side, we obtain
		\begin{align*}
		\partial_u\Theta
		&= \big(\nu \lambda \tilde{\Gamma} + \frac{\mu}{\rho}\lambda\nu \tilde{\Gamma} + r \lambda \partial_u \tilde{\Gamma} \big) r^2 (D_{vv}^2\phi)^2 + 4 \big(\nu \lambda \tilde{\Gamma} + \frac{\mu}{\rho}\lambda\nu \tilde{\Gamma} + r \lambda \partial_u \tilde{\Gamma} \big)rD_v\phi D_{vv}^2\phi \\
		&\quad + 4 \big(\nu \lambda \tilde{\Gamma} + \frac{\mu}{\rho}\lambda\nu \tilde{\Gamma} + r \lambda \partial_u \tilde{\Gamma} \big) (D_v\phi)^2 \\
		&\hspace{5em}+ \big(\frac{4 \lambda (-\nu) \mu}{\rho}\tilde{\Gamma} \big) r^2(D_{vv}^2\phi)^2 +  \frac{2\lambda(-\nu)}{\rho} \tilde{\Gamma} r^3 D_{vv}^2\phi (D_v\phi)^3 \\
		&\quad + \frac{2\lambda(-\nu)\mu}{\rho}\tilde{\Gamma}r D_{vv}^2\phi D_v\phi + \frac{8\lambda(-\nu)\mu}{\rho}\tilde{\Gamma}r D_{vv}^2\phi D_v\phi \\
		&\hspace{5em}+4 \frac{\lambda(-\nu)}{\rho}\tilde{\Gamma} r^2 (D_v\phi)^4 + 4 \frac{\lambda(-\nu)\mu}{\rho}\tilde{\Gamma}(D_v\phi)^2 \\
		&= \big(r \lambda \partial_u\tilde{\Gamma} + \frac{\lambda(-\nu)}{\rho}(4\mu-1)\tilde{\Gamma} \big) r^2 (D_{vv}^2\phi)^2 + \big(\frac{4 \nu \lambda}{\rho}\tilde{\Gamma}+\frac{10 \lambda (-\nu)\mu}{\rho}\tilde{\Gamma} + 4 r \lambda \partial_u \tilde{\Gamma} \big)\frac{r}{2 \lambda}\partial_v \big( (D_v\phi)^2 \big) \\
		&\quad + \frac{2\lambda(-\nu)}{\rho} \tilde{\Gamma} \frac{r^3}{4 \lambda}\partial_v \big((D_v\phi)^4 \big)   \\
		&\hspace{5em}+4 \frac{\lambda(-\nu)}{\rho}\tilde{\Gamma} r^2 (D_v\phi)^4 +  4 \big(\nu \lambda \tilde{\Gamma} + r \lambda \partial_u \tilde{\Gamma} \big) (D_v\phi)^2 \\
		&= - G_1 (D_{vv}^2\phi)^2 + \partial_v B - \partial_v\Big( \frac{2 \nu r }{\rho}\tilde{\Gamma} + \frac{5 (-\nu) \mu r}{\rho }\tilde{\Gamma} + 2 r^2 \partial_u \tilde{\Gamma}\Big) (D_v\phi)^2 - \partial_v\Big(\frac{(-\nu) r^3}{2 \rho}\tilde{\Gamma} \Big)(D_v\phi)^4 \\
		&\hspace{5em}+4 \frac{\lambda(-\nu)}{\rho}\tilde{\Gamma} r^2 (D_v\phi)^4 +  4 \big(\nu \lambda \tilde{\Gamma} + r \lambda \partial_u \tilde{\Gamma} \big) (D_v\phi)^2
	\end{align*}
	with $G_1$ and $B$ as stated. For the remaining terms, we compute
			\begin{align*}
		\partial_u\Theta+ G_1 (D_{vv}^2\phi)^2 
		- \partial_v B&= - \partial_v\Big( \frac{2 \nu r }{\rho}\tilde{\Gamma} + \frac{5 (-\nu) \mu r}{\rho }\tilde{\Gamma} + 2 r^2 \partial_u \tilde{\Gamma}\Big) (D_v\phi)^2 - \partial_v\Big(\frac{(-\nu) r^3}{2 \rho}\tilde{\Gamma} \Big)(D_v\phi)^4 \\
		&\hspace{5em}+4 \frac{\lambda(-\nu)}{\rho}\tilde{\Gamma} r^2 (D_v\phi)^4 +  4 \big(\nu \lambda \tilde{\Gamma} + r \lambda \partial_u \tilde{\Gamma} \big) (D_v\phi)^2 \\
		&=   - \Big(\frac{2\nu\lambda}{\rho}\tilde{\Gamma} + \frac{2\nu r}{\rho}\partial_v \tilde{\Gamma} + \frac{2 \nu \lambda r^2}{\rho}(D_v\phi)^2 \tilde{\Gamma}  - \frac{5 \nu \lambda r^2 \mu}{\rho}(D_v\phi)^2 \tilde{\Gamma} + \frac{5 (-\nu)\mu r}{\rho}\partial_v\tilde{\Gamma} \\
		&\hspace{5em}+ 5 (-\nu)\lambda r^2 (D_v\phi)^2 \tilde{\Gamma} + 4r \lambda \partial_u \tilde{\Gamma} + 2 r^2 \partial_u \partial_v \tilde{\Gamma}\Big)(D_v\phi)^2 \\
		&\hspace{5em}-\Big(\frac{-\nu \lambda r^4}{2\rho}\tilde{\Gamma}(D_v\phi)^2 + \frac{3 (-\nu)\lambda r^2}{2\rho}\tilde{\Gamma} + \frac{(-\nu)r^3}{2\rho}\partial_v \tilde{\Gamma}  \Big) (D_v\phi)^4 \\
		&\hspace{5em}+4 \frac{\lambda(-\nu)}{\rho}\tilde{\Gamma} r^2 (D_v\phi)^4 +  4 \big(\nu \lambda \tilde{\Gamma} + r \lambda \partial_u \tilde{\Gamma} \big) (D_v\phi)^2 \\
		&=  \Big( 4\nu\lambda \tilde{\Gamma} - \frac{2 \nu \lambda}{\rho}\tilde{\Gamma}  + \frac{(-\nu)(2-5\mu)}{\rho}r\partial_v \tilde{\Gamma} - 2 r^2 \partial_u \partial_v \tilde{\Gamma} \Big) (D_v\phi)^2 + \Big(\frac{\nu\lambda r^4}{2\rho}\tilde{\Gamma}  \Big) (D_v\phi)^6\\
		&\hspace{2em}+ \Big(\frac{3\nu\lambda r^2}{2\rho}\tilde{\Gamma} - \frac{2 \nu \lambda r^2}{\rho}\tilde{\Gamma} + \frac{5 \nu \lambda r^2 \mu}{\rho}\tilde{\Gamma} +5 \nu \lambda r^2 \tilde{\Gamma}+4 \frac{\lambda(-\nu)}{\rho}\tilde{\Gamma} r^2+ \frac{\nu r^3}{2\rho }\partial_v \tilde{\Gamma} \Big) (D_v\phi)^4 \\
		&=
		 \Big(\frac{2\nu\lambda(1-2\mu)}{\rho}\tilde{\Gamma} + \frac{(-\nu)(2-5\mu)}{\rho}r\partial_v \tilde{\Gamma} - 2 r^2 \partial_u \partial_v \tilde{\Gamma} \Big) (D_v\phi)^2 \\
		&\hspace{5em}+ \Big(\frac{\nu\lambda r^2}{2\rho}\tilde{\Gamma} +\frac{\nu r^3}{2\rho}\partial_v\tilde{\Gamma}  \Big)(D_v\phi)^4+ \Big(\frac{\nu\lambda r^4}{2\rho}\tilde{\Gamma}  \Big) (D_v\phi)^6 \\
		&=   - G_4 (D_v\phi)^2 - G_3 (D_v\phi)^4 - G_2 (D_v\phi)^6,
	\end{align*}
	as claimed.
\end{proof}
We first consider the unweighted identity.
\begin{lemma} \label{1/4weight}
	Let $\tilde{\Gamma}(u,v)= 1$.  Then we have
\begin{align*}
	\Theta &= r \lambda (D_{vv}^2\psi)^2,\\
	G_1 &= \frac{r^2(-\nu \lambda)(1-4\mu)}{\rho}, \\
	G_2 &= \frac{r^4 (-\nu\lambda)}{2 \rho}, \qquad G_3 = \frac{r^2 (-\nu\lambda)}{2\rho},  \\
	G_4 &= \frac{2(-\nu\lambda)(1-2\mu)}{\rho},\\
	B &= -\frac{r^3 \nu}{2 \rho}  (D_v\phi)^4 - \frac{5 r \nu \mu}{ \rho} (D_v\phi)^2 + \frac{2r\nu}{\rho}(D_v\phi)^2 .
\end{align*}
\end{lemma}
Integrating \eqref{eq:ext_prin_mainenergyestimate} over $R(-1,u)$ gives the Morawetz-type estimate
\begin{align} \label{1/4 energy}
	&\int_{\Sigma_{u}} r \lambda (\alpha')^2(u,v) dv+\iint\limits_{R(-1,u)} G_1 (D_{vv}^2\phi)^2 + G_2(D_v\phi)^6 + G_3(D_v\phi)^4 + G_4(D_v\phi)^2 dudv\\ \nonumber
	&+\int_{\underline{\Sigma}_0(-1,u)}\frac{r^3 \nu}{2 \rho}  (D_v\phi)^4  + \frac{-\nu r(2-5\mu)}{\rho}(D_v\phi)^2 du=   \int_{\Sigma_{-1}} r \lambda (\alpha')^2(-1,v) dv. 
\end{align}
\begin{remark}
	\label{rmk:ext_prin_Gamma1}
	When $\tilde{\Gamma}=1$, the coefficients that restrict coercivity are $G_1$ and $G_4$, requiring $\mu\leq\frac14$ and $\mu\leq\frac12$, respectively. In particular, this identity suffices to prove Theorem \ref{thm2} under $(\mathrm{C}_{\frac14})$.

	The estimate with $\tilde\Gamma=1$ may be viewed as a gauge-invariant nonlinear analogue of an $r^p$ estimate for the linear wave equation on Minkowski space. Indeed, let $\phi(u,v)$ solve $\partial_u\partial_v(r_0\phi)=0$, where $r_0=\frac12(v-u)$. Setting $\Theta=r_0(\partial_v^2(r_0\phi))^2$, we compute
	\begin{align}
		\partial_u \Theta &= -\frac12 (\partial_v^2(r_0\phi))^2 + 2 r_0 \partial_v^2(r_0\phi) \partial_u\partial_v^2(r_0\phi) \nonumber \\
		&= -\frac12 (\partial_v^2(r_0\phi))^2 \nonumber\\
		&= -\frac12(r_0 \partial_v^2 \phi + 2 \partial_v r_0 \partial_v\phi  +\phi \partial_v^2 r_0 )^2 \nonumber\\
		&= -\frac12 (r_0\partial_v^2\phi + \partial_v\phi)^2 \nonumber\\
		&= -\frac12( r_{0}^2 (\partial_v^2\phi)^2 + 2 r_{0}\partial_v\phi \partial_v^2 \phi + (\partial_v\phi)^2  )  \nonumber\\
		&= -\frac12 r_{0}^2 (\partial_v^2\phi)^2 - r_{0}\partial_v\phi \partial_v^2 \phi - \frac12 (\partial_v\phi)^2 \nonumber\\
		&= -\frac12 r_{0}^2 (\partial_v^2\phi)^2 - \frac12 r_{0} \partial_v(\partial_v\phi)^2  - \frac12 (\partial_v\phi)^2 \nonumber\\
		&= -\frac12 r_{0}^2 (\partial_v^2\phi)^2 - \partial_v\big(\frac12 r_{0} (\partial_v\phi)^2 \big) + \frac14 (\partial_v\phi)^2  - \frac12 (\partial_v\phi)^2 \nonumber\\
		&= -\frac12 r_{0}^2 (\partial_v^2\phi)^2 - \frac14 (\partial_v\phi)^2 - \partial_v\big(\frac12 r_{0} (\partial_v\phi)^2 \big). \label{eq:minkidentitytemp}
	\end{align}
	The nonlinear identity reduces to \eqref{eq:minkidentitytemp} after making the substitutions
	\begin{align*}
		&\hspace{3em} r \rightarrow r_0, \ \ \lambda \rightarrow \frac12, \ \ \nu \rightarrow -\frac12, \\
		&D_v\phi \rightarrow 2 \partial_v \phi, \ \ D_{vv}^2\phi \rightarrow 4 \partial_v^2 \phi, \ \  \alpha' \rightarrow 4 \partial_v^2(r_0 \phi), 
	\end{align*}
	and setting the ``purely nonlinear terms'' $\mu, (D_v\phi)^4, (D_v\phi)^6$ to zero.
\end{remark}
We now introduce a self-similar weight into Proposition \ref{prop:energy}.
\begin{lemma}\label{lem:multiplier_s_sigma}
	Let $\tilde{\Gamma}(u,v)= x^\sigma$ for some $\sigma \in (0,1)$.  Then we have
	\begin{align*}
		\Theta &= r \lambda x^\sigma  (D_{vv}^2\psi)^2,\\
		G_1 &= \frac{r^2 x^\sigma (-\nu\lambda)}{\rho}(1-4\mu+\rho\sigma+\sigma x \frac{\rho}{\nu}), \\
		G_2 &= \frac{r^4 x^\sigma(-\nu\lambda)}{2 \rho}, \qquad G_3 = \frac{r^2 x^\sigma (-\nu\lambda)}{2 \rho}(1+\sigma),  \\
		G_4 &= \frac{(-\nu\lambda) x^\sigma}{\rho}(2(1-\sigma^2)+\mu(2\sigma^2+\sigma-4)-2\sigma^2 x \frac{\rho}{\nu}),\\
		B &= -\frac{r^3 x^\sigma \nu}{2 \rho}  (D_v\phi)^4 + \frac{r(-\nu)}{\rho} x^\sigma (D_v\phi)^2  (5\mu-2-2\sigma\rho-2x\frac{\sigma\rho}{\nu}).
	\end{align*}
\end{lemma}
\begin{proof}
	First, we compute the derivatives of the self-similar weight:
	\begin{align*}
		\partial_u \tilde{\Gamma} &= \sigma x^{\sigma-1} \big( \frac{\nu}{|u|} + \frac{r}{|u|^2} \big) \\
		&= \sigma x^{\sigma-1}|u|^{-1}(\nu + x) \\
		&= \sigma x^{\sigma} r^{-1} (\nu + x),\\
		\partial_v \tilde{\Gamma} &= \sigma x^{\sigma-1} \lambda |u|^{-1}  \\
		&= \sigma x^{\sigma} \lambda r^{-1},\\
		\partial_u \partial_v \tilde{\Gamma} &= \partial_v \big(\sigma x^{\sigma} r^{-1} (\nu + x) \big) \\ 
		&= \sigma^2 x^{\sigma-1} r^{-1}(\nu+x) \partial_v x - \sigma x^{\sigma} r^{-2}\lambda (\nu+x) + \sigma x^{\sigma} r^{-1} \partial_v (\nu + x) \\
		&= \sigma^{2}x^{\sigma-1}r^{-1}(\nu+x)\lambda |u|^{-1} - \sigma x^{\sigma} r^{-2} \lambda (\nu +x) + \sigma x^{\sigma} r^{-1} \Big( \frac{\mu \lambda \nu}{\rho r} + \lambda |u|^{-1}  \Big)\\
		&= (\sigma^2 - \sigma) x^{\sigma}r^{-2} \lambda (\nu  +x) + \sigma x^{\sigma} \mu \rho^{-1} r^{-2} \lambda \nu + \sigma x^{\sigma} \lambda r^{-1}|u|^{-1} \\
		&= \big( (\sigma^2 - \sigma + \sigma \mu \rho^{-1})\nu + \sigma^2 x \big) x^{\sigma} r^{-2}\lambda.
	\end{align*}
	Substitution into Proposition \ref{prop:energy} gives the coefficients and flux
	\begin{align*}
		G_1 &= \Big( -r\big(\sigma x^{\sigma} r^{-1} (\nu + x) \big ) + \frac{(-\nu)}{\rho}x^{\sigma}(1-4 \mu)  \Big) r^2 \lambda \\ 
		&= \Big( \sigma\big((-\nu) - x \big) + \frac{(-\nu)}{\rho}(1-4\mu)   \Big )x^{\sigma}r^2 \lambda \\
		&= \Big( (-\nu)(1-4\mu + \rho \sigma ) - \sigma \rho x   \Big) x^{\sigma} \rho^{-1} r^2 \lambda \\
		&= \big( 1- 4\mu + \rho \sigma - \sigma \rho (-\nu)^{-1}x \big) \frac{\lambda(-\nu)}{\rho} x^{\sigma}r^2, \\
		G_2 &= \frac{(-\nu)\lambda}{2\rho} x^{\sigma} r^4 , \\
		G_3 &= \frac{(-\nu)\lambda}{2\rho} r^2 x^{\sigma} + \frac{(-\nu)}{2\rho}r^3 \sigma x^{\sigma} \lambda r^{-1} \\
		&= \frac{(-\nu)\lambda}{2\rho}(1+\sigma)x^{\sigma} r^2, \\
		G_4 &= \frac{2(-\nu)\lambda}{\rho}(1-2 \mu) x^\sigma - \sigma \frac{(-\nu)\lambda}{\rho}(2- 5\mu) x^{\sigma} + 2 r^2 \big( (\sigma^2 - \sigma + \sigma \mu \rho^{-1})\nu + \sigma^2 x \big) x^{\sigma} r^{-2}\lambda \\
		&=  \big( (-\nu)\rho^{-1}( 2-4\mu - \sigma(2-5\mu) - 2(\sigma^2\rho-\sigma \rho + \sigma \mu )) + 2 \sigma^2 x  \big)\lambda x^{\sigma} \\
		&= \big( 2-4\mu - \sigma(2-5\mu) - 2(\sigma^2\rho-\sigma \rho + \sigma \mu ) + 2 \sigma^2 (-\nu)^{-1}\rho x   \big)\frac{\lambda(-\nu)}{\rho}x^{\sigma}\\
		&= \big( 2(1-\sigma^2) + \mu (2 \sigma^2 + \sigma - 4) + 2 \sigma^2 (-\nu)^{-1}\rho x   \big)\frac{\lambda(-\nu)}{\rho}x^{\sigma}, \\
		B &= \frac{(-\nu)}{2\rho} x^{\sigma}r^3 (D_v\phi)^4 + \frac{5\mu (-\nu)}{\rho} x^{\sigma} r (D_v\phi)^2 + \frac{2 \nu}{\rho} x^{\sigma} r (D_v\phi)^2 + 2 r^2 (\sigma x^{\sigma} r^{-1} (\nu + x)) (D_v\phi)^2 \\
		&= \frac{(-\nu)}{2\rho} x^{\sigma}r^3 (D_v\phi)^4 + \big( 5\mu -2 - 2 \sigma \rho + 2 \sigma \rho (-\nu)^{-1} x \big)\frac{(-\nu)}{\rho} x^{\sigma}r (D_v\phi)^2.
	\end{align*}
\end{proof}
Integrating \eqref{eq:ext_prin_mainenergyestimate} over $R(-1,u)$ yields
\begin{align} \label{3/8 energy}
	&\int_{\Sigma_{u}} r \lambda x^\sigma (\alpha')^2(u,v) dv+\iint\limits_{R(-1,u)} G_1 (D_{vv}^2\phi)^2 + G_2(D_v\phi)^6 + G_3(D_v\phi)^4 + G_4(D_v\phi)^2 dudv\\ \nonumber
	&-\int_{\underline{\Sigma}_0(-1,u)}B_0(u) du=   \int_{\Sigma_{-1}} r \lambda x^\sigma (\alpha')^2(-1,v) dv. 
\end{align}

We conclude by analyzing the signs of the coefficients in Lemma \ref{lem:multiplier_s_sigma}. Since $\Theta\geq0$, coercivity of \eqref{3/8 energy} follows if $G_i\geq0$ for $1\leq i\leq4$. Global positivity is not needed; it suffices to establish positivity in an interior self-similar region
$\mathcal{B}_{\epsilon} \doteq \{ \frac{r}{|u|} \leq \epsilon \}$,
for some $\epsilon>0$ sufficiently small.

\begin{itemize}
    \item The coefficients $G_2$ and $G_3$ are non-negative for $\sigma\geq0$.
    \item $G_1$ is non-negative precisely when 
    \begin{equation*}
        1 - 4\mu + \rho \sigma - \sigma \rho (-\nu)^{-1} x \geq 0
    \end{equation*}
    holds. In $\mathcal{B}_{\epsilon}$ the term proportional to $x$ is of size $O(\epsilon)$, and one can check that the remaining terms are non-negative for $\mu \in [0, \mu_{crit,1}(\sigma)]$, where
    \begin{equation*}
        \mu_{crit,1}(\sigma) \doteq \frac{1+\sigma}{4+\sigma}.
    \end{equation*}
    Setting $\sigma = 0$ recovers the bound $\mu \leq \frac14$.
    \item $G_4$ is non-negative precisely when 
    \begin{equation*}
        2(1-\sigma^2)+\mu(2\sigma^2 + \sigma - 4) + 2\sigma^2 (-\nu)^{-1}\rho x \geq 0 
    \end{equation*}
    holds. Again, the final term is $O(\epsilon)$ in $\mathcal{B}_{\epsilon}$, and the remaining terms are non-negative for $\mu \in [0, \mu_{crit,2}(\sigma)]$, where 
    \begin{equation*}
        \mu_{crit,2}(\sigma) \doteq \frac{2(1-\sigma^2)}{4 -\sigma - 2 \sigma^2}.
    \end{equation*}
\end{itemize}

\begin{figure}[htbp]
    \centering
\includegraphics[width=.7\textwidth]{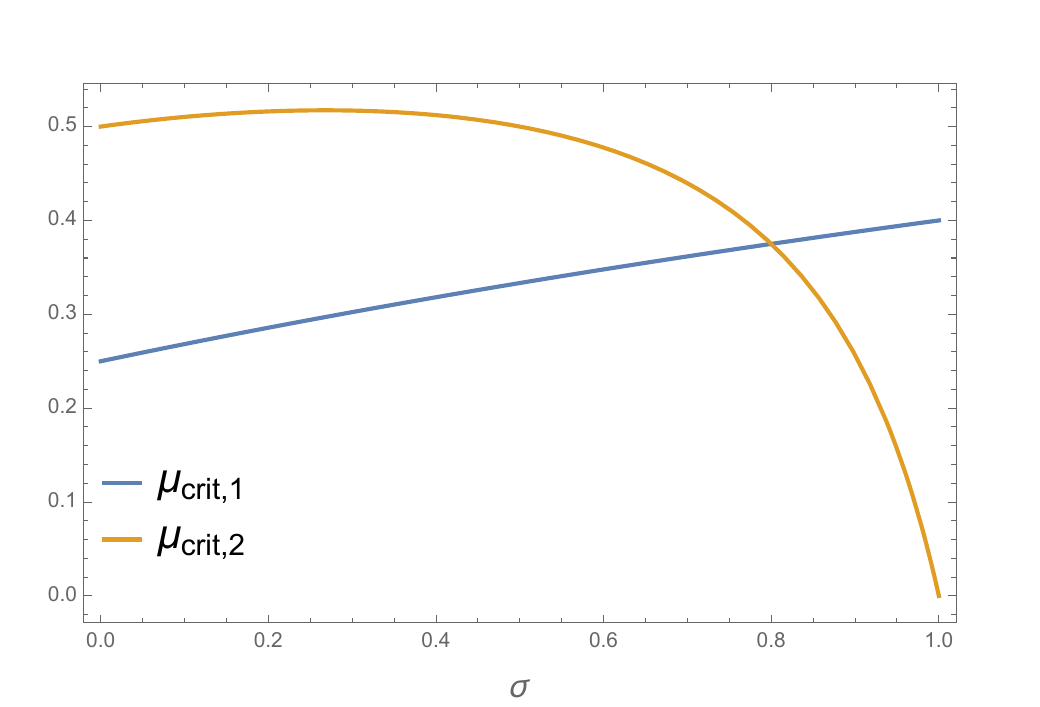}
    \caption{Graphs of $\mu_{crit,1}(\sigma)$ and $\mu_{crit,2}(\sigma)$.}
    \label{fig:mucrit1_mucrit2}
\end{figure}

The two critical curves in Figure \ref{fig:mucrit1_mucrit2} have a unique intersection at
\[
    (\sigma,\mu)=\left(\frac45,\frac38\right).
\]
Thus the optimal choice is $\sigma=\frac45$, which yields coercivity under the global condition $\mu<\frac38$, upon restriction to a sufficiently narrow interior self-similar region.

\newpage
\section{Proof of non-concentration of $\mu$ for $\mu\leq\frac14$} \label{mu less 1/4}
We now prove Theorem \ref{thm2}. By Theorem \ref{thm1}, it suffices to show that $\mu(u,v)\to0$ uniformly as $(u,v)\to\mathcal O$. Equivalently, we must rule out an $\epsilon>0$ and a sequence $(u_i,v_i)\to(0,0)$ with $u_i\nearrow0$ such that $\mu(u_i,v_i)\geq\epsilon$ for every $i$.

Assume that such a concentrating sequence exists. By \eqref{eq:3.9}, these points cannot remain in $\mathcal{C}_\delta$ for any fixed $\delta>0$; thus concentration must occur at a scale faster than the self-similar scale. To rule this out, we use the higher-order energy estimate. For $\tilde{\Gamma}=1$, integrate the identity in Proposition \ref{prop:energy} over $R(u_1,u_2)$ (see \eqref{1/4 energy}) for arbitrary $-1\leq u_1<u_2<0$, and use $(\mathrm{C}_{\frac14})$ to obtain
\begin{align}
	\label{eq:temp5}
	\iint_{R(u_1,u_2)}(\partial_v\phi)^2 (-\nu) dudv \lesssim 1 + \int_{\underline{\Sigma}_0(u_1,u_2)} \Big(-\frac{r^3 \nu}{2\lambda^4 \rho}  (\partial_v\phi)^4 - \frac{5 r \nu \mu}{\lambda^2 \rho} (\partial_v\phi)^2 + \frac{2r\nu}{\lambda^2\rho}(\partial_v\phi)^2 \Big)du.
\end{align}
To estimate the boundary terms in \eqref{eq:temp5}, note that \eqref{eq:3.6} implies
\begin{align*}
	h(u) \doteq -\frac{r^3 \nu}{2\lambda^4 \rho}  (\partial_v\phi)^4 + \frac{2 r \nu}{\lambda^2 \rho}(\partial_v\phi)^2 < 0
\end{align*}
for all sufficiently small $|u|$. Thus, after choosing $|u_1|$ sufficiently small, it remains only to estimate
\begin{align*}
	\int_{\underline{\Sigma}_0(u_1,u_2)} \frac{5 r (-\nu) \mu}{\lambda^2 \rho} (\partial_v\phi)^2 du &\lesssim \int_{\underline{\Sigma}_0(u_1,u_2)} \frac{ (-\nu) \mu}{\rho r}  du \lesssim 1,
\end{align*}
where we used \emph{(F)}. Hence the left-hand side of \eqref{eq:temp5} is finite.

A related bulk integral involving the Hawking mass is controlled by \eqref{temp2.5}. Integrating that identity over $R(u_1,u_2)$ gives
\begin{align}
	\label{eq:temp6}
	\iint_{R(u_1,u_2)}\frac{2\lambda(-\nu)\mu}{\rho r^2}dudv \leq \iint_{R(u_1,u_2)}\frac{(-\nu)}{\lambda \rho}(\partial_v\phi)^2 dudv + \int_{\underline{\Sigma}_0(u_1,u_2)}\frac{(-\nu)\mu}{\rho r}du \lesssim 1.
\end{align}
Here we used $\mu=O(r^2)$ to omit the boundary term at $r=0$. Since the estimate is uniform in $u_1$ and $u_2$, the bulk integral is finite over all of $\mathcal Q$.

\vspace{1em}
Return to the sequence $(u_i,v_i)$ with $\mu(u_i,v_i)\geq\epsilon$, and set $r_i=r(u_i,v_i)$. Choose $u_{*,i}$ and $v_{*,i}$ so that
\[
    r(u_i,v_{*,i})-r_i=r(u_{*,i},v_i)-r_i=r_i,
\]
and define the characteristic rectangle
\[
    \mathcal R_i=[u_{*,i},u_i]\times[v_i,v_{*,i}].
\]
We claim that, after passing to a subsequence, the rectangles $\mathcal R_i$ are pairwise disjoint. Suppose that $(u_{i_k},v_{i_k})$, $1\leq k\leq N$, have been chosen so that $r_{i_k}$ is strictly decreasing in $k$.

First, there exists a constant $c$, independent of $i$, such that $\sup_{\mathcal R_i}r\leq cr_i$. The supremum is attained at $(u_{*,i},v_{*,i})$. Since $\lambda$ is bounded above and below by positive constants, $v_{*,i}-v_i\sim r_i$; since $-\nu$ is bounded below, $u_i-u_{*,i}\lesssim r_i$. Moreover, because $\nu$ is decreasing in $v$,
\[
    r(u_{*,i},v)-r(u_i,v)
    \geq r(u_{*,i},v_i)-r(u_i,v_i)=r_i.
\]

We compute 
\begin{align*}
	r(u_{*,i},v_{*,i})-r(u_{*,i},v_i) & = \int_{v_i}^{v_{*,i}}\lambda(u_{*,i},v')\,dv' \\
	&\lesssim  v_{*,i}-v_i \\ 
	&\lesssim r_i.
\end{align*}
Since $r(u_{*,i},v_i) = 2r_i$, we obtain the claimed bound for $r(u,v)$ in $\mathcal{R}_i$.

Since $(u_i,v_i)\to(0,0)$, we may choose $(u_l,v_l)$, with $l>i_N$, so that $cr_l<r_{i_N}$. On $\mathcal R_l$ we have $r\leq cr_l$, whereas on each previously selected $\mathcal R_{i_k}$ we have $r\geq r_{i_k}\geq r_{i_N}$. Thus $\mathcal R_l$ is disjoint from every $\mathcal R_{i_k}$, $1\leq k\leq N$. This inductively produces the desired subsequence.

We henceforth work with this subsequence, so the $\mathcal R_i$ are pairwise disjoint. By construction, $r|_{\mathcal R_i}\sim r_i$ and $m|_{\mathcal R_i}\gtrsim\mu_i r_i\gtrsim\epsilon r_i$, so $\mu|_{\mathcal R_i}\gtrsim\epsilon$.

Estimating the integral in \eqref{eq:temp6} over $\mathcal{Q}$ gives
\begin{align*}
	\iint_{\mathcal{Q}}\frac{m}{r^3}(-\nu)dudv &\geq \sum_{i=1}^{\infty}\iint_{\mathcal{R}_i}\frac{m}{r^3}(-\nu)dudv \\
	&\gtrsim \epsilon \sum_{i=1}^{\infty}\iint_{\mathcal{R}_i}\frac{1}{r^2}(-\nu)dudv \\
	&\gtrsim \epsilon \sum_{i=1}^{\infty} r_i^{-2} \iint_{\mathcal{R}_i}(-\nu)dudv\\
	&\gtrsim \epsilon  \sum_{i=1}^{\infty} r_i^{-2} \int_{v_i}^{v_{*,i}}\big( r(u_{*,i},v)-r(u_i,v)\big)dv \\
	&\gtrsim  \epsilon  \sum_{i=1}^{\infty} r_i^{-1}\int_{v_i}^{v_{*,i}} dv\\
	&\gtrsim  \epsilon  \sum_{i=1}^{\infty} 1 \\
	&=\infty.
\end{align*}
This contradicts \eqref{eq:temp6}. Hence $\mu$ does not concentrate at the endpoint $\mathcal{O}$ and instead becomes uniformly small there. Christodoulou's extension theorem, Theorem \ref{thm1}, therefore applies and proves Theorem \ref{thm2}.

\section{BV estimates and the proof for $\mu<\frac38$} \label{sec:mu less 3/8}
{We now prove Theorem \ref{thm3}.} The key is to combine the exterior self-similar estimates of Section \ref{sec:selfsimilar} with the energy identity for the self-similar weight from Section \ref{subsec:extprin_multipliers} to obtain a global weighted estimate.

Suppose that $(\mathcal{Q},g,r,\phi)$ is a $C^1$-admissible interior satisfying the standing assumptions and the global bound
\begin{equation*}
    \sup_{\mathcal{Q}} \mu(u,v) \leq \frac38 - \delta,
\end{equation*}
for some fixed $\delta>0$. Set $\sigma=\sigma_*\doteq\frac45$ and choose $\epsilon>0$, depending on $\delta$, sufficiently small that all coefficients $\{G_i\}_{i=1}^4$ in Lemma \ref{lem:multiplier_s_sigma} are strictly positive in $\mathcal B_\epsilon=\{x\leq\epsilon\}$. We fix this $\epsilon$ and suppress the dependence of the implicit constants on it.

\begin{proposition}
    \label{prop:ext_prin_selfsimilar_multiplier}
    In the exterior self-similar region\footnote{Recall that $\mathcal{C}_\epsilon=\mathcal{Q}\setminus\mathcal{B}_\epsilon$ lies to the exterior of the self-similar curve $\{r/|u|=\epsilon\}$.} $\mathcal{C}_\epsilon$, we have
    \begin{equation}
        \label{eq:ext_prin_selfsimilar_multiplier}
        \iint_{\mathcal{C}_\epsilon} r^2 (D_{vv}^2\phi)^2(-\nu)dudv \lesssim 1.
    \end{equation}
\end{proposition}
\begin{proof}
    Let $\chi_\epsilon:[0,\infty)\to[0,1]$ be a non-negative increasing cutoff satisfying $\chi_\epsilon(y)=0$ for $y\leq\frac12\epsilon$ and $\chi_\epsilon(y)=1$ for $y\geq\epsilon$.

    By \eqref{eq:3.4}, we may decrease $\epsilon$ so that $r(u,0)/|u|>\epsilon$. Thus $\tilde\Gamma=1$ and $\partial_u\tilde\Gamma=0$ on $\{v=0\}$, so the boundary flux there contains no cutoff-derivative term.

    Apply \eqref{eq:ext_prin_mainenergyestimate} over $R(u_1,u_2)$, where $u_1<u_2$, with multiplier $\tilde\Gamma=\chi_\epsilon(r/|u|)$. Derivatives of $\tilde\Gamma$ are supported where $r/|u|\in[\frac12\epsilon,\epsilon]$. In particular,
    \begin{align*}
        \partial_u \tilde{\Gamma} &= \chi_\epsilon'\big(\frac{r}{|u|}\big)\big( \frac{\nu}{|u|} + \frac{r}{|u|^2} \big)  \\
        &= \chi_\epsilon'\big(\frac{r}{|u|}\big) |u|^{-1} (\nu + \frac{r}{|u|}).
    \end{align*}
    Since $r/|u|=O(\epsilon)$ on the support of $\chi_\epsilon'$ and $-\nu$ is bounded below by a positive constant independent of $\epsilon$, choosing $\epsilon$ sufficiently small ensures that $\partial_u\tilde\Gamma\leq0$. Its contribution to $G_1$ therefore has the favorable sign.

    Moreover, on the support of $\tilde\Gamma$ we are exterior to a fixed self-similar cone, so $\mu\to0$ uniformly by \eqref{eq:3.9}. After taking $u_1$ sufficiently close to zero, we therefore have $\mu<\frac14$ on $R(u_1,u_2)\cap\operatorname{supp}\tilde\Gamma$ for every $u_2\in(u_1,0)$. The cutoff contributions to $G_3$ and $G_4$ are supported in the transition region and can be controlled by the exterior estimates.

    Integrating \eqref{eq:ext_prin_mainenergyestimate} over $R(u_1,u_2)$ gives
    \begin{align*}
       \int_{\{u=u_2\}}& \chi_\epsilon  \lambda r (\alpha')^2 dv +  \iint_{R(u_1,u_2)} \frac{(-\nu)\lambda}{2\rho} r^4 \chi_\epsilon (D_v\phi)^6 dudv   \\
       &\hspace{-4em}+ \iint_{R(u_1,u_2)}\Big( r\lambda (-\partial_u \chi_\epsilon) + \chi_\epsilon \frac{(-\nu)\lambda(1-4\mu)}{\rho} \Big) r^2 (D_{vv}^2\phi)^2 dudv\\
       &= \int_{\{u=u_1\}} \chi_\epsilon  \lambda r (\alpha')^2 dv   + \int_{ \{v=0, u_1 \leq u \leq u_2\} } \Big( -\frac{r^3 \nu}{2\rho}(D_v\phi)^4 - \frac{5 r \nu \mu}{\rho}(D_v\phi)^2 + \frac{2 r \nu}{\rho } (D_v\phi)^2  \Big)\chi_\epsilon du \\
       &+ \iint_{R(u_1,u_2)}\Big(\frac{r^2 \nu \lambda}{2\rho} \chi_\epsilon + \frac{r^3 \nu}{2\rho}\partial_v \chi_\epsilon \Big)(D_v\phi)^4 dudv \\
       &+ \iint_{R(u_1,u_2)}\Big( \frac{2\nu\lambda(1-2\mu)}{\rho}\chi_\epsilon + \frac{r(-\nu)(2-5\mu)}{\rho}\partial_v \chi_\epsilon - 2 r^2 \partial_u\partial_v \chi_\epsilon \Big)(D_v\phi)^2 dudv.
    \end{align*}
    It follows that
    \begin{align*}
        \iint_{\mathcal{C}_\epsilon\cap R(u_1,u_2)} &r^2 (D_{vv}^2\phi)^2 (-\nu)dudv \\
        &\lesssim \int_{\{u=u_1\}} \chi_\epsilon  \lambda r (\alpha')^2 dv   + \int_{ \{v=0, u_1 \leq u \leq u_2\} } \Big( -\frac{r^3 \nu}{2\rho}(D_v\phi)^4 - \frac{5 r \nu \mu}{\rho}(D_v\phi)^2 + \frac{2 r \nu}{\rho } (D_v\phi)^2  \Big) du \\
        &+ \iint_{\mathcal{C}_{\epsilon/2}\cap R(u_1,u_2)} \Big(r^2 (D_v\phi)^4  + (D_v\phi)^2 \Big)(-\nu)dudv \lesssim 1,
    \end{align*}
    Here we used \eqref{eq:3.6} to see that the boundary term on $\{v=0\}$ is non-positive for sufficiently small $|u_1|$, the pointwise estimate \eqref{eq:3.10} for $r\partial_v\phi$, and the integrated exterior estimate \eqref{eq:3.12}, applied with $\delta=\epsilon/2$. The bound is uniform in $u_2$, so letting $u_2\to0$ controls the portion with $u\geq u_1$. The remaining portion lies in a compact regular region and has finite integral, proving the proposition.
\end{proof}

We next propagate this integrated control from the exterior region to the axis.
\begin{proposition}
    \label{prop:ext_prin_axis_multiplier}
In the region $\mathcal{B}_\epsilon$ we have 
\begin{equation}
    \iint_{\mathcal{B}_\epsilon} \Big(\frac{r}{|u|}\Big)^{\sigma_*} r^2 (D_{vv}^2\phi)^2(-\nu)dudv + \iint_{\mathcal{B}_\epsilon} \Big(\frac{r}{|u|}\Big)^{\sigma_*}  (D_v\phi)^2(-\nu)dudv  \lesssim 1.
\end{equation}
\end{proposition}
\begin{proof}
    Let $\tilde\chi_\epsilon:[0,\infty)\to[0,1]$ be a non-negative decreasing cutoff satisfying $\tilde\chi_\epsilon(y)=1$ for $y\leq\epsilon$ and $\tilde\chi_\epsilon(y)=0$ for $y\geq\frac32\epsilon$.

    We apply \eqref{eq:ext_prin_mainenergyestimate} with $\tilde{\Gamma}=\tilde{\chi}_\epsilon(x)x^{\sigma_*}$. The coefficients of the bulk terms associated with $\tilde{\Gamma}=x^{\sigma_*}$ were computed in Lemma \ref{lem:multiplier_s_sigma}. As shown there, if $\mu\leq\frac38-\delta$ and $x\leq\epsilon\ll1$ with $\epsilon$ sufficiently small, then all four coefficients $\{G_i\}_{i=1}^4$ are strictly positive.

    Derivatives of $\tilde\chi_\epsilon$ produce error terms supported where $r/|u|\in[\epsilon,\frac32\epsilon]\subset\mathcal C_\epsilon$. These terms are controlled by \eqref{eq:3.12} and \eqref{eq:ext_prin_selfsimilar_multiplier}.

    The rest of the proof follows the argument of Proposition \ref{prop:ext_prin_selfsimilar_multiplier}.
\end{proof}

Combining the interior and exterior estimates yields the global bound
\begin{equation}
    \label{eq:ext_prin_globalenergy}
    \iint_{\mathcal{Q}} \Big(\frac{r}{|u|}\Big)^{\sigma_*}r^2 (D_{vv}^2\phi)^2 (-\nu) dudv + \iint_{\mathcal{Q}} \Big(\frac{r}{|u|}\Big)^{\sigma_*} (D_v\phi)^2 (-\nu) dudv \lesssim 1.
\end{equation}

Define the BV norm of the induced data on the outgoing cone $\Sigma_u$ by
\begin{equation*}
    \text{BV}(u) \doteq \int_{\Sigma_u} |\alpha'(u,v) | \lambda(u,v) dv,
\end{equation*}
and set
\begin{equation*}
    \mathcal{E}(u) \doteq \int_{\Sigma_u} \Big(r^{2+\sigma_*}(D_{vv}^2\phi)^2 + r^{\sigma_*}(D_v\phi)^2 \Big)dv.
\end{equation*}

By Cauchy--Schwarz,
\begin{align*}
    \text{BV}(u) &\lesssim \int_{\Sigma_u} r|D_{vv}^2\phi| \lambda dv +  \int_{\Sigma_u} |D_v\phi| \lambda dv \\
    &\lesssim \Big( \int_{\Sigma_u}  r^{-\sigma_*} dv\Big)^{\frac12}\Big( \int_{\Sigma_u} r^{2+\sigma_*}(D_{vv}^2\phi)^2 dv \Big)^{\frac12} + \Big(\int_{\Sigma_u} r^{-\sigma_*}dv \Big)^{\frac12}\Big(\int_{\Sigma_u} r^{\sigma_*} (D_v\phi)^2 dv \Big)^{\frac12}\\
    &\lesssim |u|^{\frac12 - \frac12 \sigma_*} \mathcal{E}(u)^{\frac12}.
\end{align*}
It follows that 
\begin{align}
    \label{eq:ext_prin_BVintegral}
    \int_{-1}^{0}\frac{\text{BV}(u)^2}{|u|} du &\lesssim \int_{-1}^{0}\frac{\mathcal{E}(u)}{|u|^{\sigma_*}}du \lesssim 1,
\end{align}
where the final inequality follows from \eqref{eq:ext_prin_globalenergy}. Thus the BV norm decays along a sequence: there exists $u_i\nearrow0$ such that $\mathrm{BV}(u_i)\to0$.

For sufficiently large $i$, Christodoulou's small-BV theory \cite{christ2} gives a uniformly small BV variation throughout the future domain of dependence of $\Sigma_{u_i}$. Since $rD_v\phi=\alpha-\phi$, integration by parts and the radial mass equation then show that $\mu$ is uniformly small in this region. Christodoulou's extension theorem, Theorem \ref{thm1}, therefore yields the $C^1$-extension property and completes the proof of Theorem \ref{thm3}.

\providecommand{\MR}{\relax\ifhmode\unskip\space\fi MR }
\providecommand{\MRhref}[2]{%
  \href{http://www.ams.org/mathscinet-getitem?mr=#1}{#2}
}
\providecommand{\href}[2]{#2}

\end{document}